\documentclass[pdflatex,sn-mathphys-ay]{sn-jnl}% Math and Physical Sciences Author Year Reference Style
\usepackage{graphicx}%
\usepackage{multirow}%
\usepackage{amsmath,amssymb,amsfonts}%
\usepackage{amsthm}%
\usepackage{mathrsfs}%
\usepackage[title]{appendix}%
\usepackage{xcolor}%
\usepackage{textcomp}%
\usepackage{manyfoot}%
\usepackage{booktabs}%
\usepackage{algorithm}%
\usepackage{algorithmicx}%
\usepackage{algpseudocode}%
\usepackage{listings}%
\usepackage{aasmacros}
\usepackage{natbib}
\usepackage{caption}
\usepackage{tabularx}
\theoremstyle{thmstyleone}%
\theoremstyle{thmstyletwo}%

\theoremstyle{thmstylethree}%

\begin{document}

\title[]{\centering Isotopic Evidence for a Cold and Distant Origin of the Interstellar Object 3I/ATLAS\\ {\normalsize \textit{Preprint --- In Review at Nature: March $6^{th}$ 2026}}}


\author*[1,2]{\fnm{Martin} \sur{Cordiner}}\email{martin.cordiner@nasa.gov}

\author[1,3]{\fnm{Nathan X.} \sur{Roth}}

\author[4]{\fnm{Marco} \sur{Micheli}}

\author[1]{\fnm{Geronimo} \sur{Villanueva}}

\author[5]{Davide Farnocchia}

\author[1]{\fnm{Steven} \sur{Charnley}}

\author[6]{Nicolas Biver}

\author[6]{Dominique Bockel{\'e}e-Morvan}

\author[7]{\fnm{Dennis} \sur{Bodewits}}

\author[8,9,10]{Colin Orion Chandler}

\author[6]{Jacques Crovisier}

\author[11]{\fnm{Maria N.} \sur{Drozdovskaya}}

\author[12]{\fnm{Kenji} \sur{Furuya}}

\author[13]{\fnm{Michael S. P.} \sur{Kelley}}

\author[1]{\fnm{Stefanie} \sur{Milam}}

\author[7]{\fnm{John W.} \sur{Noonan}}

\author[14]{Cyrielle Opitom}

\author[15]{Megan E. Schwamb}

\author[16]{Cristina A. Thomas}

\affil*[1]{\orgdiv{Solar System Exploration Division}, \orgname{NASA Goddard Space Flight Center}, \orgaddress{\street{8800 Greenbelt Road}, \city{Greenbelt}, \postcode{MD 20771}, \country{USA}}}

\affil[2]{\orgdiv{Department of Physics}, \orgname{Catholic University of America}, \orgaddress{\street{620 Michigan Ave. NE}, \city{Washington}, \postcode{DC 20064}, \country{USA}}}

\affil[3]{\orgdiv{Department of Physics}, \orgname{American University}, \orgaddress{\street{4400 Massachusetts Ave. NW}, \city{Washington}, \postcode{DC 20016}, \country{USA}}}

\affil[4]{\orgdiv{ESA NEO Coordination Centre}, \orgname{Planetary Defence Office, European Space Agency}, \orgaddress{\street{Largo Galileo Galilei, 1}, \postcode{00044},  \city{Frascati}, \state{RM}, \country{Italy}}}

\affil[5]{Jet Propulsion Laboratory, California Institute of Technology, 4800 Oak Grove Dr., Pasadena, CA 91109, USA}

\affil[6]{LIRA, Observatoire de Paris, Universit\'e PSL, CNRS, Sorbonne Universit\'e, Universit\'e Paris Cit\'e, CY Cergy Paris Universit\'e, 5 place Jules Janssen, F-92190 Meudon, France}

\affil[7]{\orgdiv{Department of Physics}, \orgname{Auburn University}, \orgaddress{\street{Edmund C. Leach Science Center}, \postcode{36382},  \city{Auburn}, \state{AL}, \country{USA}}}

\affil[8]{Dept. of Astronomy \& the DiRAC Institute, University of Washington, 3910 15th Ave NE, Seattle, WA 98195, USA}
\affil[9]{LSST Interdisciplinary Network for Collaboration and Computing, 933 N. Cherry Avenue, Tucson, AZ 85721, USA}
\affil[10]{Dept. of Astronomy \& Planetary Science, Northern Arizona University, PO Box 6010, Flagstaff, AZ 86011, USA}

\affil[11]{\orgname{Physikalisch-Meteorologisches Observatorium Davos und Weltstrahlungszentrum (PMOD/WRC)}, \orgaddress{Dorfstrasse 33}, \postcode{7260} \city{Davos Dorf}, \country{Switzerland}}

\affil[12]{\orgdiv{Pioneering Research Institute}, \orgname{RIKEN}, \orgaddress{2-1 Hirosawa}, \city{Wako-shi, Saitama}, \postcode{351-0198}, \country{Japan}}

\affil[13]{\orgdiv{Department of Astronomy}, \orgname{University of Maryland}, \orgaddress{\street{4296 Stadium Dr., PSC Bldg. 415}, \postcode{20742-2421},  \city{College Park}, \state{MD}, \country{USA}}}

\affil[14]{Institute for Astronomy, University of Edinburgh, Royal Observatory, Edinburgh EH9 3HJ, UK}

\affil[15]{Astrophysics Research Centre, School of Mathematics and Physics, Queen's University Belfast, Belfast BT7 1NN, UK}

\affil[16]{Northern Arizona University, Department of Astronomy and Planetary Science, P.O. Box 6010, Flagstaff, AZ, 86011 USA}


\abstract{Interstellar objects provide the only directly observable samples of icy planetesimals formed around other stars, and can therefore provide insight into the diversity of physical and chemical conditions occurring during exoplanet formation. Here we report isotopic measurements of the interstellar comet 3I/ATLAS, which reveal an elemental composition unlike any Solar System body. The water in 3I/ATLAS is enriched in deuterium, at a level of D/H = ($0.95\pm0.06$)\%, which is more than an order of magnitude higher than in known comets, while its range of $^{12}$C/$^{13}$C ratios (141--191 for CO$_2$ and 123--172 for CO) exceeds typical values found in the Solar System, as well as nearby interstellar clouds and protoplanetary disks. Such extreme isotopic signatures indicate formation at temperatures $\lesssim30$~K in a relatively metal-poor environment, early in the history of the Galaxy. When interpreted with respect to models for Galactic chemical evolution, the carbon isotopic composition implies that 3I/ATLAS accreted roughly 10--12 billion years ago, following an early period of intense star formation. 3I/ATLAS thus represents a preserved fragment of an ancient planetary system, and provides direct evidence for active ice chemistry and volatile-rich planetesimal formation in the young Milky Way.}


\maketitle

\section{Introduction}\label{sec:intro}

Interstellar objects can carry materials formed around other stars to within a close enough range of Earth for detailed spectroscopic observations. The idea that a large number of small, icy bodies populate the interstellar medium (ISM) of our Galaxy has been discussed for decades \citep{whi75,mcg89,coo16}. The presence of such ``interstellar comets'' is consistent with our understanding of how planetesimals accrete and are ejected by gravitational interactions during the formation of planetary systems \citep{don04,lev10,zhe25}. The discoveries of the interstellar objects (ISOs) 1I/`Oumuamua, 2I/Borisov and 3I/ATLAS on hyperbolic trajectories through our Solar System \citep{jew23,sel25}, provide startling validation of these theories. Considering the difficulty of studying the ices in the midplanes of protoplanetary disks and planetary systems elsewhere in our Galaxy \citep{stu24}, spectroscopic observations of ISOs now provide revolutionary insights for investigating how, and from which materials, planetesimals formed in systems other than our own. 

Isotopic ratio measurements provide sensitive probes of interstellar physical and chemical conditions \citep{nom23}, and thus have the ability to place constraints on the nature of the interstellar object's originating environment. In contrast to $^{13}$C, which becomes enriched in the ISM over Galactic timescales, the deuterium (heavy hydrogen) elemental abundance becomes depleted, as a result of stellar processing \citep{gal95,dea96}. However, the D/H ratios in interstellar molecules and (pre-)planetary materials often deviate significantly from this trend. Deuterium enrichments reach several orders of magnitude in prestellar and protostellar environments, where ion-molecule chemistry is the main driver of deuteration in the cold gas \citep{nom23}. Consequently, the D/H ratio in water ice provides a measure of the environmental ionisation conditions and H$_2$O formation temperature \citep{taq14,lee15,jen21}. 

The exact point of origin of 3I/ATLAS (in space and time), and the history of its path through the Galaxy, are presently highly uncertain \citep{hop25,tay25,guo25}. A broad range of ages (3--11) Gyr was deduced based on its high velocity, but tracing the orbital trajectory back more than $\sim10$~Myr is difficult due to unpredictable (chaotic) gravitational interactions within the inhomogeneous Galactic environment \citep{per25}. Additional, isotopic information will help illuminate 3I/ATLAS's origin, leading to a better appreciation of how it can fit into, and help improve our understanding of, the processes of planet formation and Galactic chemical evolution.

Here we present James Webb Space Telescope (JWST) NIRSpec spectral imaging of the interstellar comet 3I/ATLAS during the outbound leg of its 2025 perihelion passage. We detected emission lines from gas-phase H$_2$O, CO$_2$, and CO molecules (major isotopologues), in addition to their respective minor isotopologues HDO, $^{13}$CO$_2$ and $^{13}$CO. The HDO/H$_2$O ratio provides new insights into the physical and chemical conditions in which the constituent ices of 3I/ATLAS formed and accreted, while the $^{12}$C/$^{13}$C ratio reveals the degree of heavy-element enrichment (metallicity) of the gas, leading to new insights into its Galactic age and place of origin.

\section{Observations and Results}\label{sec:results}

Observations of 3I/ATLAS were performed using JWST in the near-infrared range, covering the H$_2$O fundamental rovibrational bands (at 2.7~$\mu$m) on 2025-Dec-22, with a subsequent, deeper integration covering H$_2$O (hot bands), HDO, CO$_2$, $^{13}$CO$_2$, CO and $^{13}$CO (between 3.6--5.0~$\mu$m) on 2025-December-23 (see Methods for details). A summary of the observed molecular bands is given in Table \ref{tab:mols}.  Complementary, ground-based submillimeter-wave observations of 3I/ATLAS's CO and HCN emission were obtained using the Atacama Compact Array (ACA) on 2025-Dec-22 (see Methods), for the purpose of measuring the coma expansion velocity. 3I/ATLAS was at a heliocentric distance $r_H=2.4$~au for all observations.  

\begin{figure*}
\begin{center}
\includegraphics[height=6.5cm]{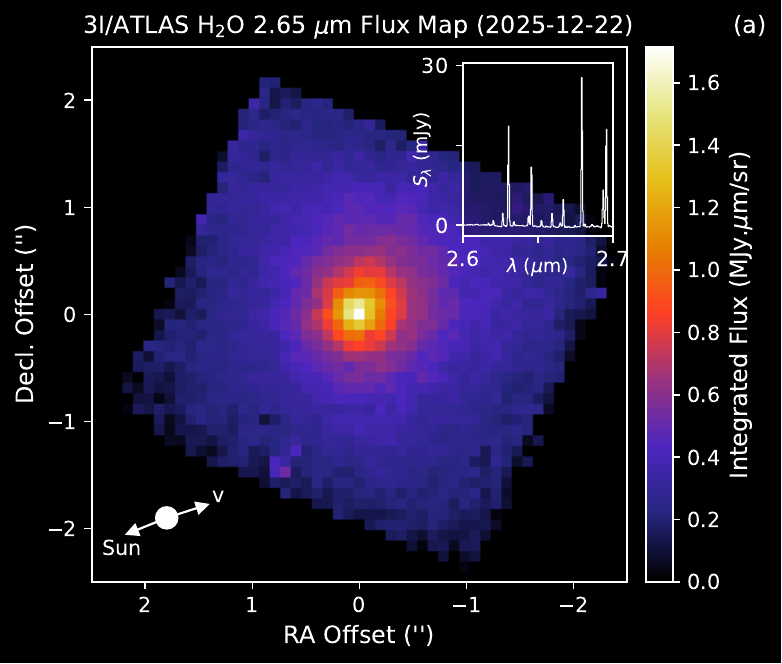}
\includegraphics[height=6.5cm]{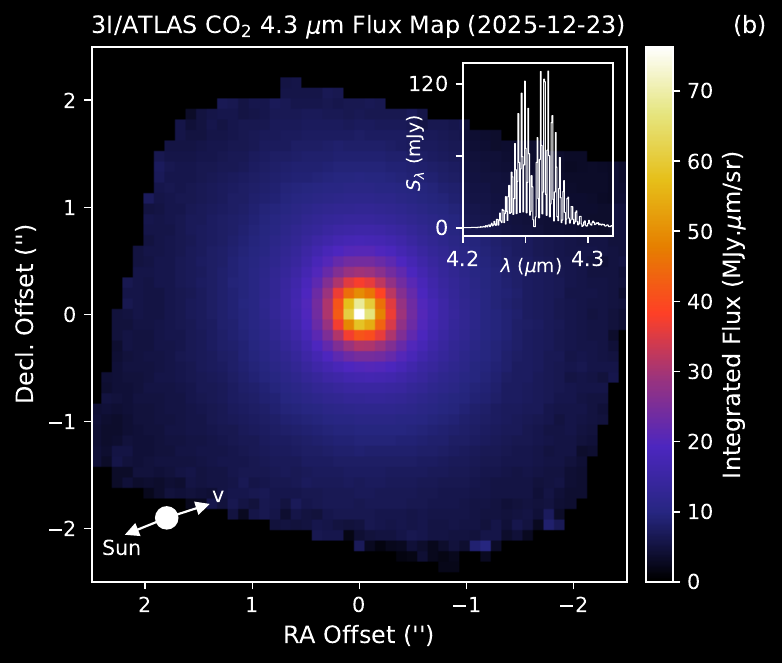}
\includegraphics[height=6.5cm]{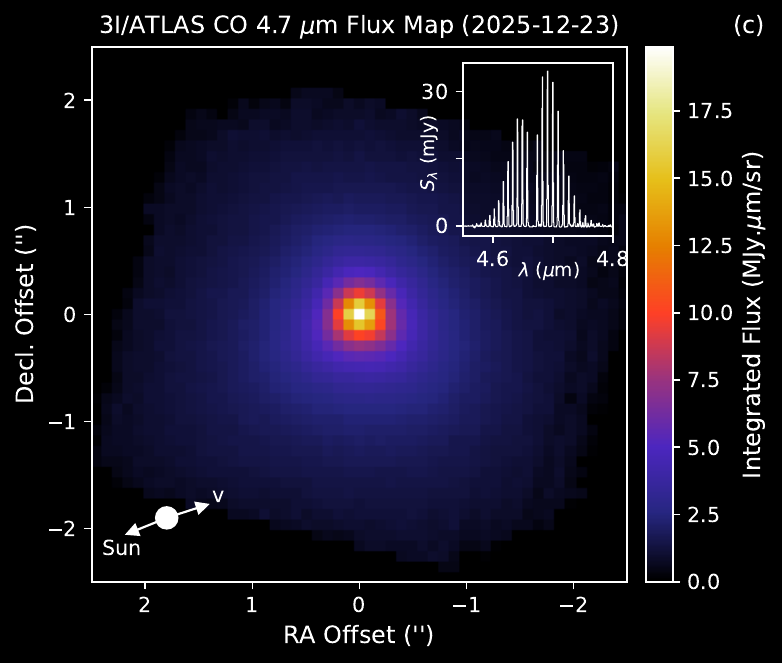}
\end{center}
\caption{Spectrally integrated line flux maps for 3I/ATLAS observed using JWST NIRSpec: (a) H$_2$O at 2.7~$\mu$m, (b) CO$_2$ at 4.3~$\mu$m, (c) CO at 4.7~$\mu$m. Image axes are aligned with the equatorial (RA/decl.) grid. Molecular emission has been isolated by subtracting a polynomial fit to the adjacent continuum. Spatial coordinates are with respect to the brightest pixel in each map (the pseudo-nucleus). Inset panels (upper right) show the continuum-subtracted spectra within the image integration wavelength range, spatially integrated within a $3''$ diameter circular aperture centered on the pseudo-nucleus (corresponding to 3,900 km at the 1.8~au distance of the comet from the telescope). Lower left corner shows the direction of the (sky-projected) comet-Sun (S) and nucleus velocity ($v$) vectors. }
\label{fig:maps}
\end{figure*}

\begin{figure*}
\begin{center}
\includegraphics[width=0.33\textwidth]{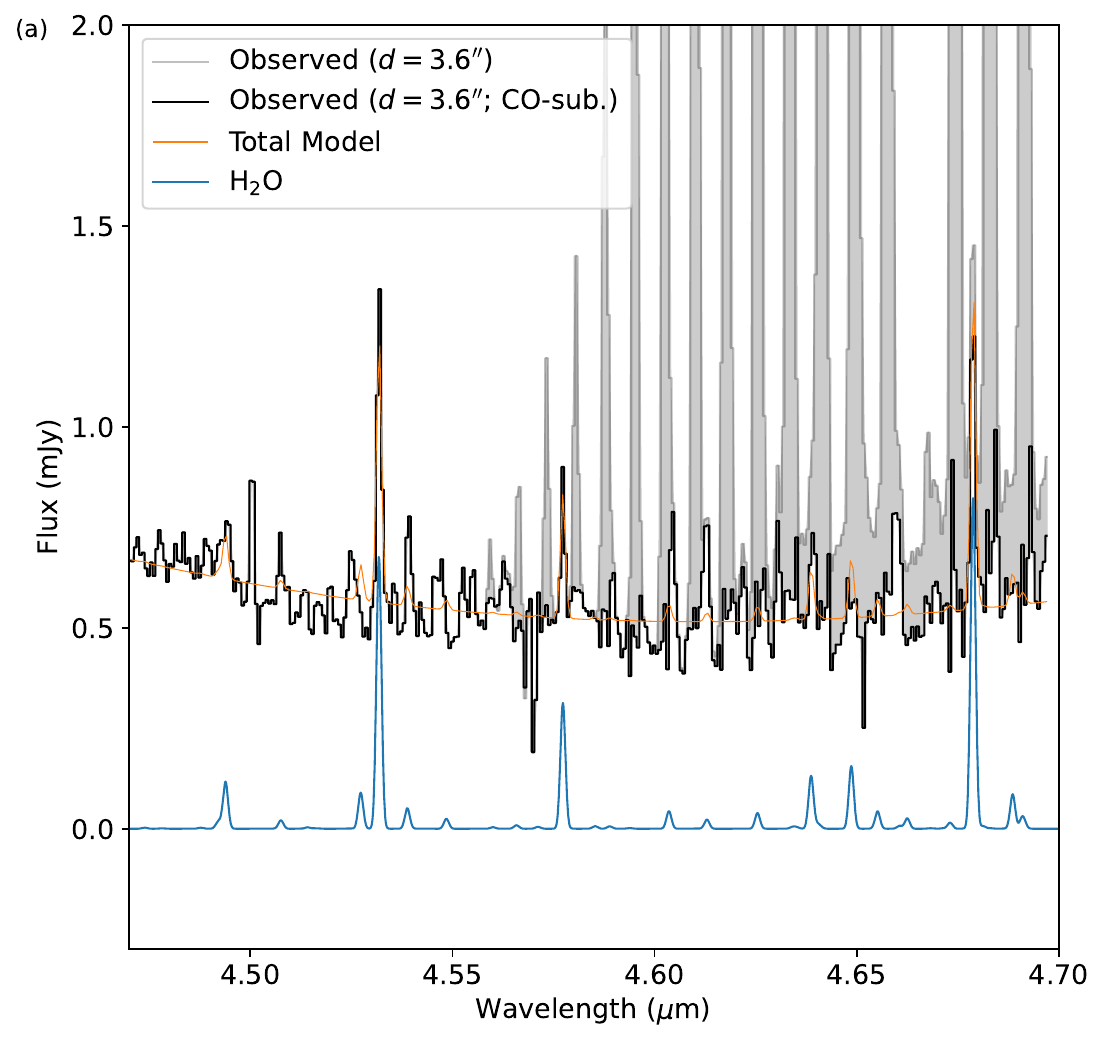}
\includegraphics[width=0.32\textwidth]{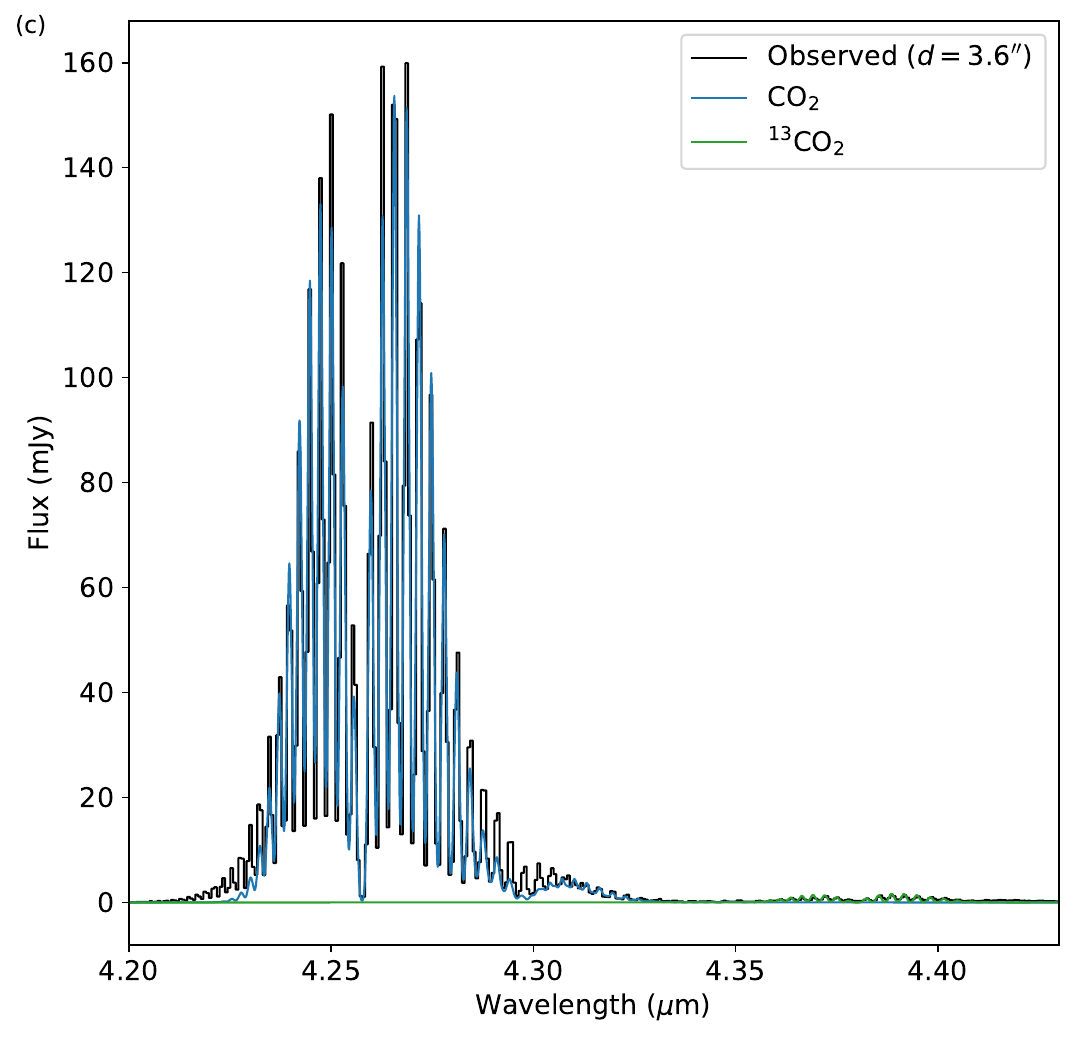}
\includegraphics[width=0.33\textwidth]{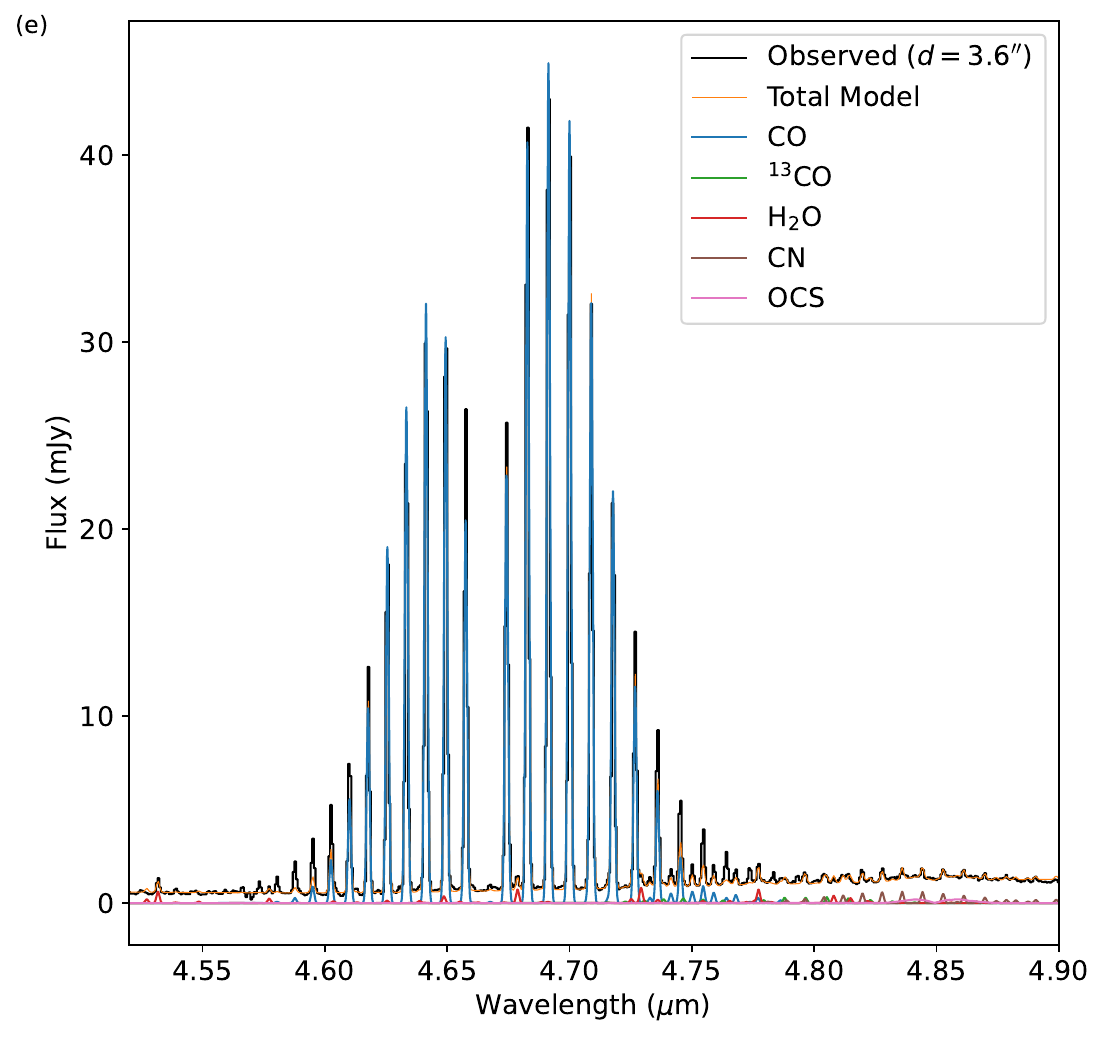}
\includegraphics[width=0.328\textwidth]{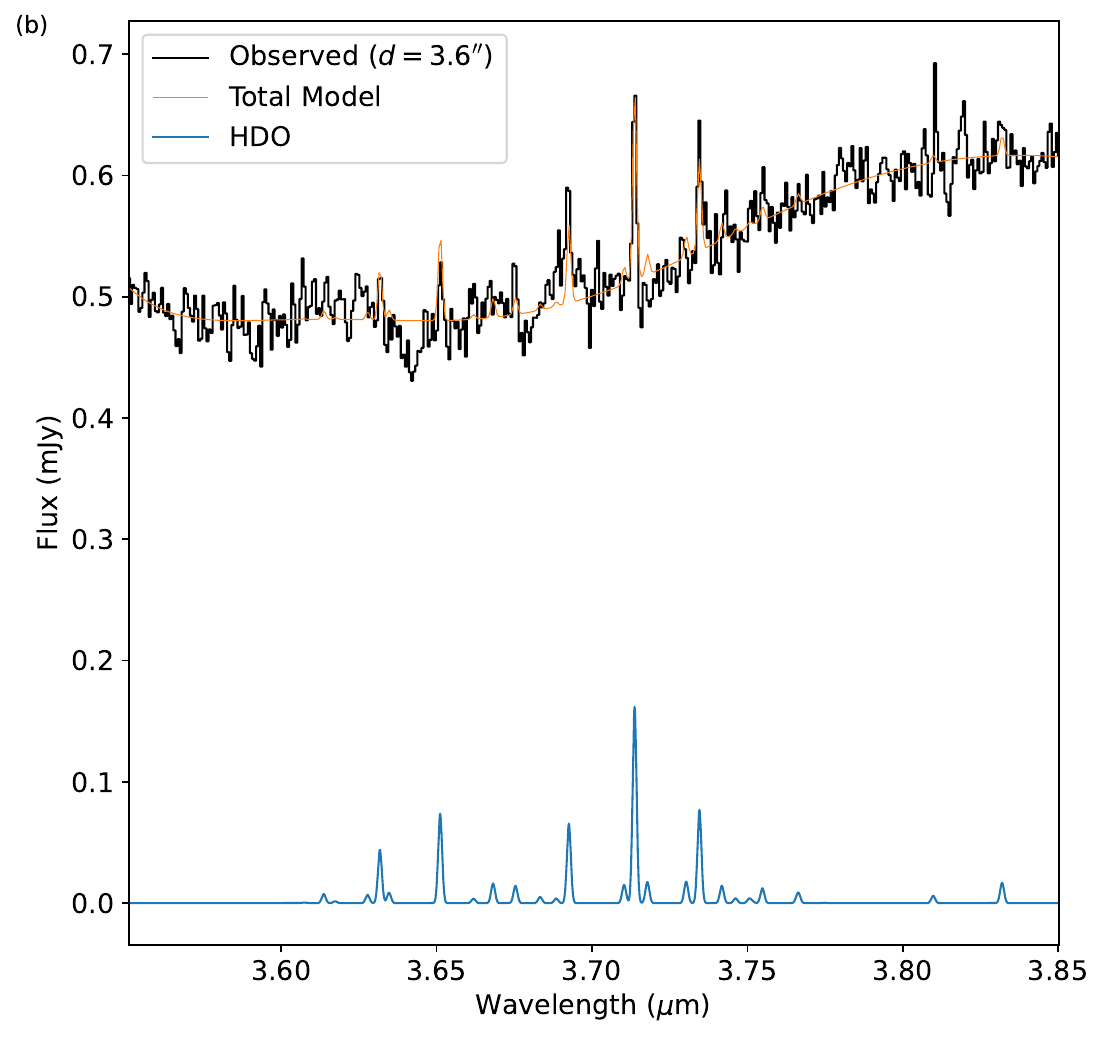}
\includegraphics[width=0.328\textwidth]{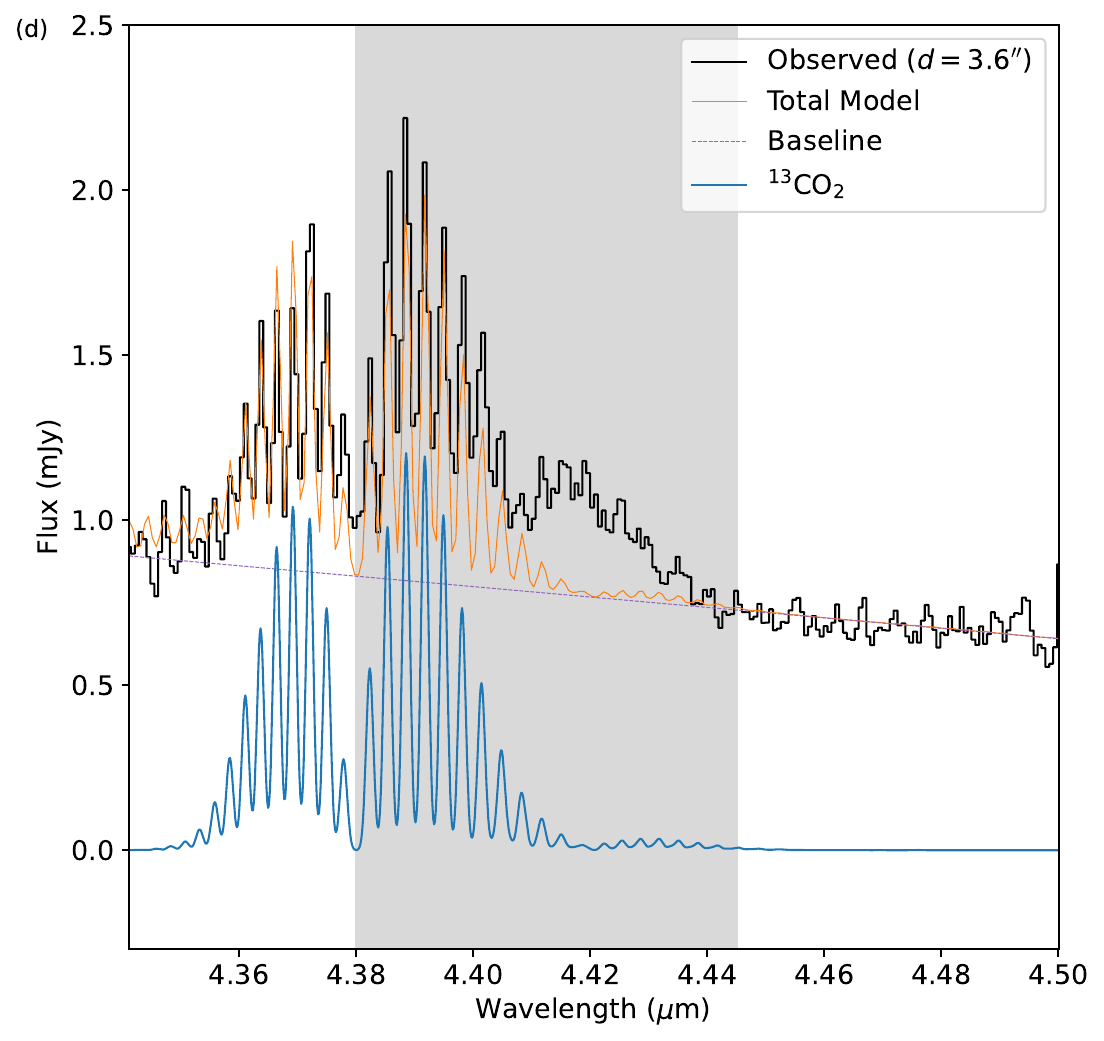}
\includegraphics[width=0.328\textwidth]{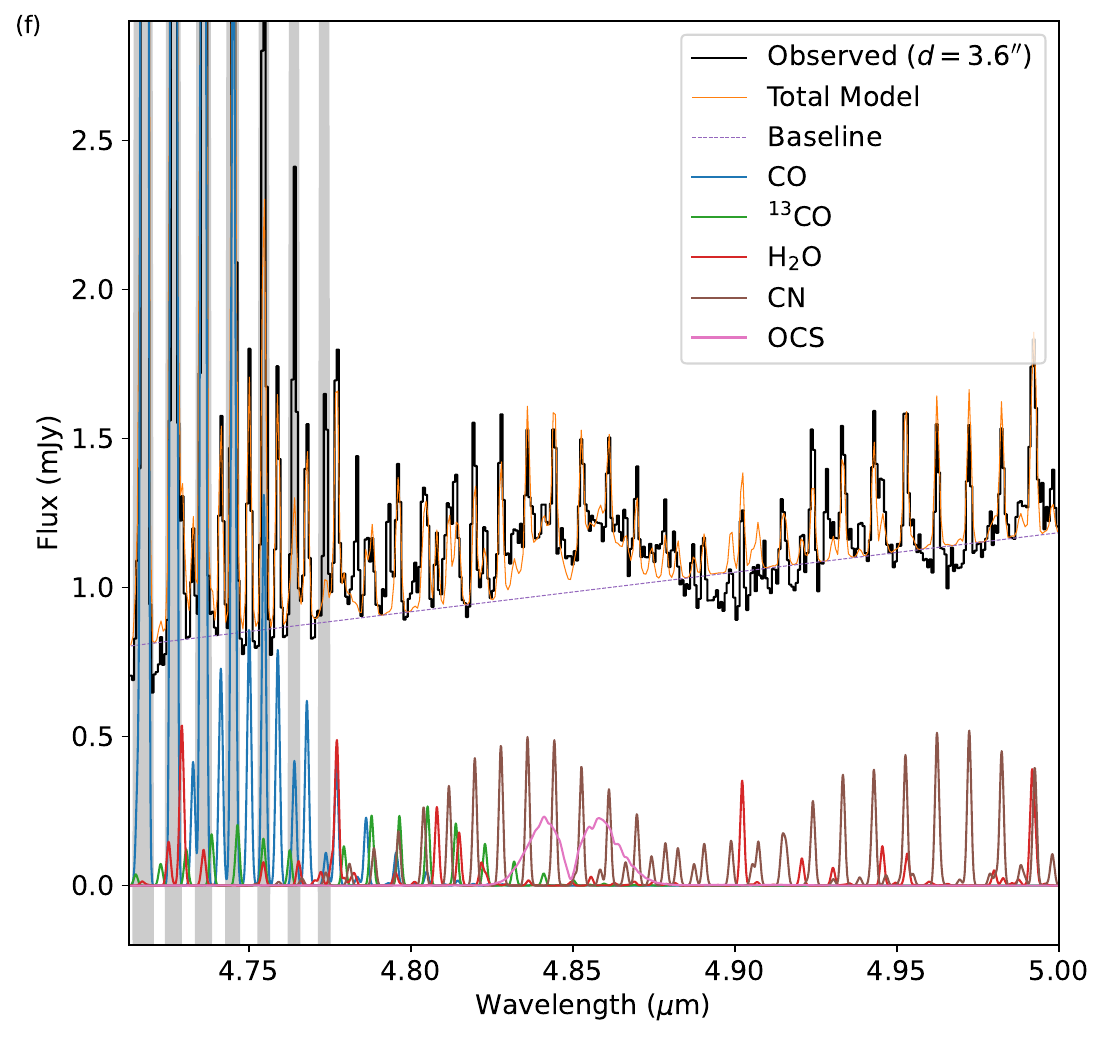 }
\end{center}
\caption{Observed NIRSpec molecular spectra of 3I/ATLAS integrated within a $d=3.6''$-diameter circular aperture centered on the pseudo-nucleus. Similarly spatially-integrated, best-fitting spectral models are overlaid. As shown by the grey filled regions in panel (a), the strong CO ($v=1-0$) lines were subtracted prior to fitting, to reduce spectral contamination when fitting H$_2$O. The grey region in panel (d) indicates wavelengths excluded (masked) from the fitting due to potential contamination by an unidentified feature around 4.42~$\mu$m. In panel (f), the regions surrounding the stronger CO lines (shown in grey) were masked during $^{13}$CO fitting, to avoid contamination by the CO model residuals, and to allow improved fitting of the higher-$J$ CO lines, some of which overlap $^{13}$CO. For panels (d) and (f), first-order baseline fits are shown; see Extended Data Fig. \ref{fig:bases} for the ensemble of baselines used in deriving our final range of $^{12}$C/$^{13}$C values for these species.}
\label{fig:fits}
\end{figure*}

Multiple spectral lines from each of the major isotopologues were detected across the entirety of the NIRSpec field of view, confirming the presence of a spatially extended gas coma, with a peak intensity at the assumed position of the nucleus (the pseudo-nucleus). The image cubes for the major isotopologues were baseline-subtracted using a polynomial fit to the surrounding continuum regions in each pixel, then spectrally integrated to produce the molecular emission maps shown in Fig. \ref{fig:maps}. Spatially-integrated spectra (within a $d=3.6''$-diameter circular aperture centered on the pseudo-nucleus), are shown for all the species of interest in Fig. \ref{fig:fits}. 

Gas production rates ($Q$), rotational temperatures ($T_{rot}$), and isotopic ratios were derived by modeling the observed infrared emission lines as a function of sky-projected nucleocentric distance ($\rho$) using the Planetary Spectrum Generator (PSG; \citealt{villanueva18}) (see Methods). The resulting $Q_{\rho}$ curves for CO$_2$ and CO  (Extended Data Fig. \ref{fig:qcurves}) reach a clear asymptote (``terminal'' $Q$ value) at $\rho\sim0.3''$ ($\sim400$~km), indicating a lack of significant gas production for these species at larger radii, consistent with the majority of their outgassing arising directly from the nucleus. H$_2$O shows a significant increase in $Q_{\rho}$ between $\rho\sim0.3''$--$1.7''$ ($\sim500$--2,200~km), consistent with a contribution to the H$_2$O gas production from icy grain sublimation in the coma; uniformity of the HDO/H$_2$O ratio with radius (within the errors), implies similar gas release mechanisms for both H$_2$O and HDO.  Terminal gas production rates ($Q_t$) were calculated for each species from the average of the $Q_{\rho}$ values in the 5 outermost annuli shown in Extended Data Fig. \ref{fig:qcurves}, and are given in Table \ref{tab:mols}. From these production rates, we find the following mixing ratios for the main coma gas constituents within the JWST field of view: CO$_2$/H$_2$O = $1.04\pm0.03$; CO/H$_2$O = $2.33\pm0.07$; CO/CO$_2$ = $2.24\pm0.01$. Since the 2025-Aug-06 JWST observations of \citet{cor25} (obtained at $r_H =3.32$~au inbound), the coma CO$_2$/H$_2$O ratio has fallen by a factor of $\approx7$, while the CO/H$_2$O and CO/CO$_2$ ratios have increased by factors of $\approx1.4$ and $\approx10$, respectively. This corresponds to a transition from a CO$_2$-dominated to a CO-dominated coma during the object's passage past the Sun, indicating a remarkable temporal evolution of the relative outgassing rates for these coma gases.

\begin{table}
\caption{\label{tab:mols} 3I/ATLAS Observed Gas Production Rates}
\centering
\begin{tabularx}{10cm}{XXXX}
\hline\hline
Gas & Band ID & Wavelegth & $Q_t$\\
    &         & ($\mu$) & ($10^{25}$~s$^{-1}$)\\
\hline
H$_2$O& $\nu_1$ + $\nu_3$ & 2.7 & $161\pm1$\\
H$_2$O& Hot bands & 4.6 & $169\pm5$\\
HDO   & $\nu_1$   & 3.6 & $3.20\pm0.19$\\
CO$_2$& $\nu_3$   & 4.3 & $176\pm1$\\
$^{13}$CO$_2$& $\nu_3$ & 4.4& 0.94--1.22\\
CO     & $v=1$--0 & 4.7 & $394\pm1$\\
$^{13}$CO& $v=1$--0& 4.8& 2.49--2.99\\
OCS   & $\nu_3$ & 4.9 & $0.47\pm0.05$\\
CN    & $\nu_3$ & 4.9 & $>0.31$\\
\hline
\end{tabularx}
\parbox{10cm}{Notes --- Gas production rates ($Q_t$) in molecules per second, are averaged between $\rho=0.8$--$1.8''$ to mitigate opacity effects from the inner coma, apart from $^{13}$CO, which was averaged between $\rho=0.4$--$1.4''$ due to its weaker lines (see Methods). Errors on $Q_t$ are the $1\sigma$ statistical model-fitting uncertainties. For $^{13}$CO and $^{13}$CO$_2$, the $Q_t$ ranges correspond to to the ensemble of best-fitting values across a range of plausible baseline models. }
\end{table}

From the simultaneously-measured $Q_t$(HDO) and $Q_t$(H$_2$O) values (see Methods), we derive a D/H ratio for water of $(0.95\pm0.06)$\%.  Baseline definition is more challenging for $^{13}$CO$_2$ and $^{13}$CO than HDO, so for those species we report the results considering an ensemble of best-fitting $Q_t$ values across a range of plausible baseline models (the associated data are given in Extended Data Tables \ref{tab:13co2} and \ref{tab:13co}), resulting in $^{12}$C/$^{13}$C = 141--191 for CO$_2$ and 123--172 for CO.

\section{Discussion}\label{sec:discuss}

\begin{figure*}
\begin{center}
\hspace{1mm}\includegraphics[width=0.82\textwidth]{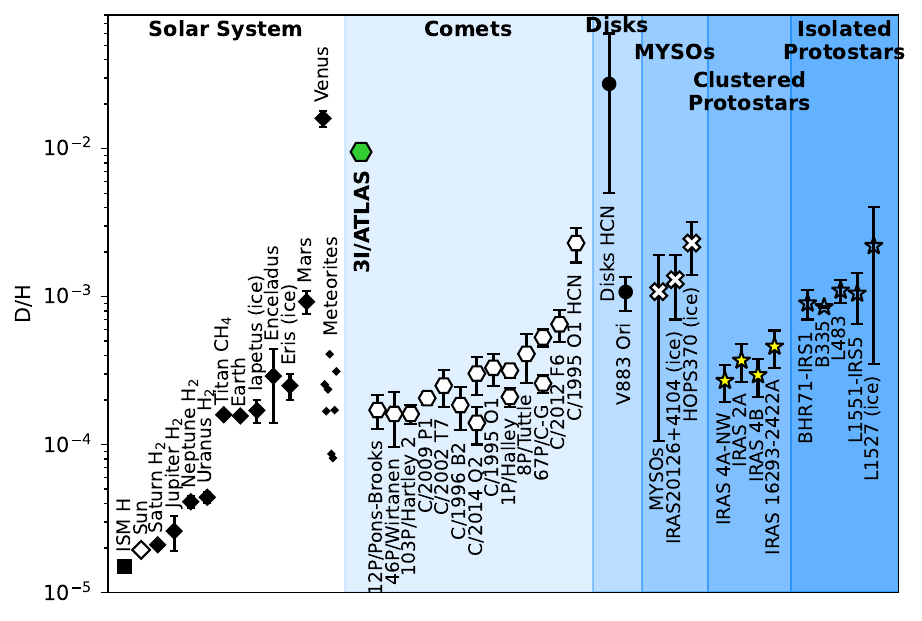}
\includegraphics[width=0.81\textwidth]{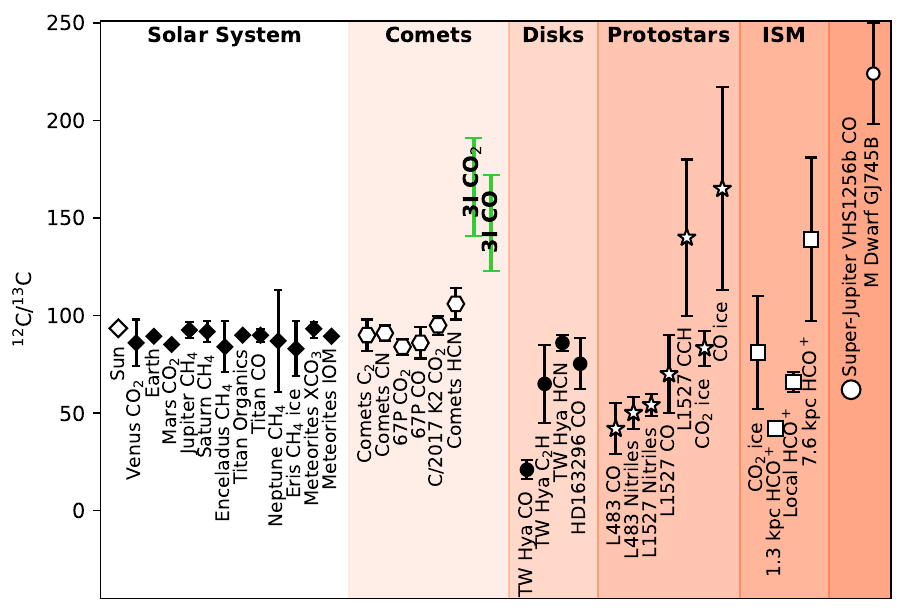}
\end{center}
\caption{Isotopic ratios observed in the coma of 3I/ATLAS compared with Galactic and Solar System observations for D/H (top) and $^{12}$C/$^{13}$C (bottom). Data are from the compilation of \citet{nom23}, with additional D/H values for V883 Ori (gas-phase; \citealt{tob23}), and IRAS20126+4104, HOPS370 and L1527 (ice-phase; \citealt{sla24,sla25}). D/H data are for H$_2$O unless otherwise specified. The ``disks HCN'' data represent the mean and range of observed D/H values across five protoplanetary disks.   For $^{12}$C/$^{13}$C, the meteoritic ``XCO$_3$'' and ``IOM'' data are the mean values (and ranges) for carbonates and insoluble organic matter, respectively, in carbonaceous chondrites \citep{ale07,ale10,ale15}; the TW Hya and HD\,163296 values for CO are from \citet{yos22,qi26}; protostellar ice CO and CO$_2$ are the mean values from \citet{bru24}; interstellar CO$_2$ ice is the average (and range) of values from \citet{boogert15}; interstellar HCO$^+$ values for three Galactocentric distances are from \citet{luo24}; super-Jupiter exoplanet CO data are from \citet{gan23}; M dwarf GJ745 data are from \citet{cro19}. Error bars are $1\sigma$ unless otherwise specified.}
\label{fig:isotopes}
\end{figure*}

The D/H and $^{12}$C/$^{13}$C ratios for 3I/ATLAS are both extremely unusual among previously observed comets and other Solar System objects (Fig. \ref{fig:isotopes}). Although previous measurements of the $^{12}$CO$_2$/$^{13}$CO$_2$ ratio have been obtained in only two comets (67P and C/2017 K2), the $^{12}$C/$^{13}$C ratio of the broader reservoir of volatile carbon in the Solar System is well characterised. Observations of CH$_4$ in Jupiter, Saturn and Neptune, CO$_2$ on Mars and Venus, CH$_4$ ice on Eris and Makemake, as well as carbonates (and insoluble organic matter; IOM) in meteorites, all show remarkably similar $^{12}$C/$^{13}$C ratios, consistent with terrestrial sedimentary rocks ($^{12}$C/$^{13}$C = 89.4), to within a few per-cent, while the Solar ratio is $93.5\pm0.7$ \citep{lyo18}. The elevated $^{12}$C/$^{13}$C ratios in 3I/ATLAS are therefore completely distinct from other materials measured in the Solar System, thus highlighting a non-local origin. Although surprising, our observed isotopic ratios are supported by the findings of Salazar Manzano et al. (submitted) and Opitom et al. (submitted), who also found elevated D/H and $^{12}$C/$^{13}$C ratios for 3I/ATLAS based on their respective ALMA and VLT observations.

Even outside the Solar System, measurements of $^{12}$C/$^{13}$C $> 100$ are rare, although such high ratios are predicted (and occasionally observed) in the outer parts of the Galaxy \citep{wil94,kob11,rom17,luo24}. The primary source of carbon in our Galaxy is from nucleosynthesis in the envelopes of asymptotic giant branch (AGB) stars, with additional, minor contributions from novae and supernovae \citep{rom03,ram14,kob11}. Over their lifetimes, intermediate-mass AGB stars return material to the interstellar medium with increasingly-elevated elemental $^{13}$C abundances, due to the conversion of $^{12}$C to $^{13}$C during hot bottom burning (\emph{via} the CN cycle). Consequently, the cycle of star formation and death leads to decreasing $^{12}$C/$^{13}$C ratios in the interstellar medium as a function of Galactic age. In particular regions where non-equilibrium chemistry can prevail, molecular $^{12}$C/$^{13}$C ratios are subject to further modulation, as a result of isotope-selective photodissociation and low-temperature chemical kinetics \citep{lan84,vis09,nom23,lee24}. These processes can either deplete or enhance the $^{12}$C/$^{13}$C ratios in hydrocarbons and nitriles, yet their impact on the $^{12}$C/$^{13}$C ratio in CO --- the main carbon reservoir in dense interstellar clouds --- is typically small relative to the $^{12}$C/$^{13}$C ratio.  The relative insensitivity of interstellar CO and CO$_2$ to isotopic fractionation processes at low temperatures contributes to the relative consistency of $^{12}$C/$^{13}$C ratios across the Solar System. Consequently, the $^{12}$C/$^{13}$C ratio in primitive planetary (and cometary) materials is thought to be close to the interstellar elemental value at the time of formation of the host star.

The elemental abundances observed in 3I/ATLAS place strict requirements on the chemical profile of its originating star-system, which, combined with dynamical constraints from the object's Galactic orbit (deduced from its inbound velocity vector), result in a set of criteria that help isolate its origin in space and time. Taken at face value, the high CO and CO$_2$ coma abundances, as well as substantial abundances of organic molecules (CH$_3$OH, H$_2$CO, CH$_4$; Roth et al., in prep.) relative to H$_2$O provide evidence for a carbon- (and oxygen-) rich originating system. Furthermore, since rocky planets and planetesimals can only form in the presence of relatively high metal abundances (characterized by a logarithmic [Fe/H] abundance $\gtrsim-0.6$, relative to the Sun; \citealt{and24,bol24}), moderately high metallicity is also required. Both these criteria can be fulfilled by an early burst of massive star formation (within a few giga-years of the Galaxy's formation), followed by a slower period of further enrichment of carbon (and other elements) by low-mass AGB stellar outflows. According to the Galactic chemical evolution model of \citet{kob11}, elevated interstellar $^{12}$C/$^{13}$C ratios $\gtrsim130$ can be achieved at relatively high metallicity ([Fe/H] $\approx-0.2$ to $-0.6$) under such a scenario. In the Galactic bulge and ``thick disk'' stellar populations, above-average metallicity and below-average $^{12}$C/$^{13}$C ratios occur as a consequence of early, rapid star formation, followed by slower $^{13}$C enrichment due to the relative lack of intermediate-mass (IM) AGB stars (with between 4--8 solar masses). Since nitrogen and $^{13}$C are both produced more efficiently in IM AGB stars, \emph{via} the CNO cycle \citep{kar14}, deficiencies of both these elements are expected to correlate. The high CO/HCN ratio of $391\pm25$ in 3I/ATLAS compared with Solar System comets (see ALMA Observations Section and \citealt{cor20}), could imply a relatively high C/N ratio, which may therefore also be consistent with the formation of 3I/ATLAS in an environment subject to early, rapid star formation, prior to the attainment of the level of $^{13}$C and N enrichment typically seen in the present-day Solar System and local Galaxy.

Although the models of \citet{kob11} indicate that high $^{12}$C/$^{13}$C ratios (up to 140) can be most readily achieved at [Fe/H] $\gtrsim-0.6$ in the Galactic bulge, the high velocity dispersion of bulge stars ($\sim150$~km\,s$^{-1}$) is inconsistent with the present orbit of 3I/ATLAS, which matches better with stars in the Galactic disk \citep{guo25,hop25}. On the other hand, the elemental composition may be consistent with the large population of old stars with elevated C/N and $\alpha$/Fe ratios, identified by \citet{mas15} in the Galactic thick disk. The Galactic trajectory of 3I/ATLAS is indeed compatible with a thick-disk origin, even if a thin-disk origin is statistically more probable (M. Hopkins 2026, private communication).

In a sample of 63 nearby solar-type stars, \citet{bot20} obtained $^{12}$C/$^{13}$C ratios in the range 71--105.  Based on their ages, which range between $\sim0$--9~Gyr, \citet{rom22} inferred a relatively rapid onset of $^{13}$C enrichment during the first 4~Gyr of our Galaxy's evolution, as a result of $^{13}$C production in novae. The resulting, steep $^{12}$C/$^{13}$C curve at early times in their model with 3--$8M_{\odot}$ white-dwarf nova progenitors (see \citealt{rom22}, Fig. 9), constrains the isotopic age of 3I/ATLAS to the range 11--12~Gyr, assuming an origin in the nearby Galactic disk. For comparison, the secondary $^{12}$C/$^{13}$C peak at [Fe/H] $=-0.5$ in \citet{kob11}'s Fig. 19 (thick-disk model), corresponds to a similar estimated age of $\sim11$ Gyr. 

Uncertainties in the modeled $^{12}$C/$^{13}$C ratios as a function of time are significant due to incomplete knowledge of the stellar initial mass function, star formation rates as a function of time (and Galactocentric radius), gas infall/outflow rates and stellar nucleosynthetic yields. A conclusive age estimate for 3I/ATLAS is further complicated by the interstellar $^{12}$C/$^{13}$C gradient as a function of Galactocentric distance ($R_{GC}$) \citep{yan23,luo24}. Although definitive measurements of interstellar gas $^{12}$C/$^{13}$C ratios comparable to the 3I/ATLAS CO$_2$ lower bound of 141 are lacking, extrapolation of the \citet{luo24} trend line for HCO$^+$ to the outer edge of the Galaxy ($R_{GC}=14$ kpc) gives $^{12}$C/$^{13}$C = $116\pm29$. Therefore, within the uncertainties of the Galactic trend, an origin for 3I/ATLAS in the most distant, metal-depleted outer regions of the Galactic disk is also a possibility. However, it remains to be demonstrated whether such an origin could be consistent with 3I/ATLAS's present-day trajectory. Furthermore, the observational trend (and models) presented by \citet{yan23} only reach $^{12}$C/$^{13}$C $\approx80$--120 in the outermost disk. In that case, significant temporal evolution would still be required to match the 3I/ATLAS $^{12}$CO$_2$/$^{13}$CO$_2$ ratio, although the age estimate would potentially be shortened to $\sim10$--11~Gyr. Ultimately, the observed range of $^{12}$C/$^{13}$ ratios in 3I/ATLAS can be most readily explained as a result of temporal evolution of the interstellar $^{12}$C/$^{13}$C ratio, with a probable additional contribution from the Galactic radial $^{12}$C/$^{13}$C gradient.

The D/H ratio for water in 3I/ATLAS demonstrates a surprisingly high deuterium enrichment compared with other (icy and non-icy) Solar System bodies and protostars within the Galaxy (Fig. \ref{fig:isotopes}). The value of $(0.95\pm0.06)$\% lies $58\sigma$ above the mean D/H of 0.029\% in Solar System comets (where $\sigma$ is the standard deviation), and $12\sigma$ above the mean water D/H of 0.1\% found in the total sample of high and low-mass protostars. Within the Solar System, only Venus's atmosphere has a higher D/H, which results from mass-dependent atmospheric escape over the planet's lifetime \citep{don82}.  By contrast, the deuteration of H$_2$O ice in Solar System comets is theorized to occur primarily in the interstellar and pre-stellar phases, where H$_2$O is rapidly synthesized on the surfaces of dust grains at very low temperatures ($\lesssim30$~K) \citep{taq14,fur17}.  The D/H ratio is there determined by the balance of gas and grain-surface reactions involving H, D and O atoms, initiated by the conversion of (unreactive) gas-phase HD into (highly reactive) H$_2$D$^+$, as a result of the reaction HD + H$_3$$^+$ $\rightleftharpoons$ H$_2$ + H$_2$D$^+$. The rate of the reverse reaction (destroying H$_2$D$^+$) is highly sensitive to both the gas kinetic temperature ($T_{kin}$) and the amount of H$_2$ in its ortho \emph{vs.} para spin states \citep{taq14,lee15}, so the resulting H$_2$O deuterium enrichments become most strongly enhanced for low values of the H$_2$ ortho/para ratio (OPR) and $T_{kin}$.

Adopting the typical metallicity, temperature and radiation conditions found in the nearby Milky Way, detailed models for the ice chemistry of interstellar clouds and protostellar envelopes are able to reproduce the gas and ice D/H values $\sim0.1$\% observed in nearby protostars and protoplanetary disks \citep{fur16,jen21}. \citet{jen21} showed that deuteration can remain efficient in star-forming regions as long as the dust temperatures remain below 30~K, while increased ultraviolet radiation fields and cosmic ray ionisation rates actually accelerate the reactions responsible for deuterating H$_2$O; a high D/H ratio $\sim1$\% is achieved in the presence of a cosmic ray flux $\sim100$ times the local Galactic value. Such enhanced-ionisation conditions would be expected in the vicinity of the lower-metallicity region of intense, massive star-formation indicated by the high $^{12}$C/$^{13}$C ratio and high abundances of $\alpha$ elements in 3I/ATLAS. A reduced CO abundance (relative to hydrogen) in low-metallicity interstellar clouds should further enhance the HDO/H$_2$O ratio (Furuya et al., in prep.), since CO acts a chemical inhibitor on the main H$_2$D$^+$ deuteration pathway \citep{pag11}.

%In the lower metallicity environment implied by our high observed $^{12}$C/$^{13}$C ratio, reduced dust grain abundances would result in less-attenuated interstellar radiation fields, and therefore higher gas ionisation rates and associated warmer dust temperatures \citep{ski11,mad16}. 

A systematic difference is observed between the HDO/H$_2$O ratios in nearby protostellar envelopes and those found in Solar System comets (Fig. \ref{fig:isotopes}). This has been suggested to originate as a result of temperature effects during ice formation and/or high temperature reprocessing of H$_2$O in the disk \citep{nom23,cor25a}. Although the number statistics are small for clustered and isolated protostars, the tendency toward lower D/H in clustered environments is theorized to be a result of higher temperatures and a shorter, low-temperature protostellar collapse phase, that leads to inhibited D enrichment \citep{jen21}. The D/H ratio in Solar System comets may be further reduced by radial mixing, or sublimation and re-freezing of water in the protoplanetary disk, resulting in the incorporation of re-equilibrated H$_2$O (with a lower D/H) back into the ice, thus lowering the bulk D/H ratio \citep{jac13}. The exceptionally high D/H ratio measured in 3I/ATLAS therefore implies H$_2$O ice formation at very cold temperatures in an irradiated interstellar or protostellar environment, followed by a lack of significant, high-temperature reprocessing of its water in the disk. Although molecular ice and planetesimal formation is expected to be less efficient in reduced metallicity environments \citep{gua22,and24}, the HDO-rich composition of 3I/ATLAS proves that the cold, ice-rich conditions required for cometary accretion can still be maintained in more strongly-irradiated, less metal-rich disks. 

%According to the chemical model of \citet{lee15} (see their Figure 6), assuming a gas density of $10^5$~cm$^{-3}$ and an equilibrium H$_2$ OPR, a D/H ratio of 0.79\% can be achieved over a timescale of $3\times10^5$~yr at $T_{kin}=20$~K

%Bayet 2010: Variations in FUV or temperature do not lead to significant changes in abundances for species such as HDO, in high-mass protostellar cores.  Molecular abundances scale according to the metallicity.

\section{Conclusion}

JWST observations of the interstellar comet 3I/ATLAS have revealed exceptionally high  D/H and $^{12}$C/$^{13}$C ratios in its coma. The observed elemental/isotopic abundance profile places strong constraints on the temperature and star-formation history of its originating environment. To obtain D/H = 0.95\%, the bulk of 3I/ATLAS's H$_2$O ice must have formed under enhanced-ionisation conditions, at temperatures $\lesssim30$~K, with minimal contamination from water (re-)processed at higher temperatures. Also taking into account the object's kinematical age estimates, a probable scenario is that 3I/ATLAS accreted as long ago as 10--12 Gyr, beyond the H$_2$O ice line, in a relatively cold, dense and well-shielded protoplanetary disk midplane. Its $^{13}$C-poor natal environment was most likely embedded in a strongly-irradiated interstellar cloud, enriched in C and O (and to lesser extents, N and S) in the aftermath of an intense period of massive star formation. The abundance of C, H, O, N and S-bearing species in 3I/ATLAS demonstrates the existence of volatile compounds around other stars, that could lead to complex, potentially pre-biotic chemistry \citep[\emph{e.g.}][]{kri23}, during the early history of our Galaxy. 

%\backmatter
\clearpage

\section*{Methods}

\subsection*{JWST Observations}

Observations of 3I/ATLAS were performed using the JWST NIRSpec integral field unit (IFU; \citealt{boker22}), as part of program ID 5094. The G235H dispersive element was used for a single 642~s exposure starting UT 2025-12-22 03:36, covering wavelengths $\lambda = 1.7$--3.2 $\mu$m, followed by $5\times700$~s exposures with the G395H disperser starting UT 2025-12-23 08:07, covering $\lambda = 2.9$--5.3 $\mu$m. The resulting $30\times30$ array of spectra have a resolving power $R_\lambda=\lambda/\Delta\lambda\sim2700$ and a pixel size of $0.1''$, which is approximately the same as the FWHM of the JWST point-spread function at 3~$\mu$m.

3I/ATLAS was acquired and tracked in the IFU using JPL Horizons ephemeris solution $\#42$. During observations, the object was 1.80~au from the telescope, at $r_H=2.37$--2.42~au, with a (Sun-target-observer) phase angle of $22.7^{\circ}$--$21.6^{\circ}$. Each exposure was divided across four dither positions, spatially separated (in the approximate shape of a square) with offsets of $\approx0.2''$ from the (central) targeted position. The data were reduced using the JWST Calibration Pipeline software version v1.20.2 \citep{bushouse25} using the JWST Calibration Reference Data System context file 1464.  The multiple dithers for each exposure were shifted and combined in the rest frame of the comet during image processing, thus allowing detector artifacts and cosmic ray events to be identified and removed. For each 3I/ATLAS exposure, a set of sky background exposures with matching exposure time, disperser and dither settings was obtained, offset by $300''$ from the science target, along the horizontal axis of the IFU aperture. Unfortunately, two of the five background exposures for G395H failed due to guide-star acquisition issues; these exposures will be re-observed later in 2026. Inspection of the available background exposures showed no evidence for significant emission from any background (\emph{e.g.} interstellar) infrared sources or zodiacal light, so the final target data cubes were reduced without background subtraction in order to maximize the signal-to-noise ratio (SNR). 

After centering on the comet, the individual exposures were combined and rebinned onto a common spatial-spectral coordinate system using the 3D drizzle algorithm included as part of the JWST pipeline's {\tt cube\_build} task  \citep{law23}. The resulting image cube spatial dimensions were $\approx3.7''\times3.5''$ for G235H and $\approx4.4''\times4.0''$ for G395H. The absolute flux calibration accuracy of the final data cubes is expected to be 3\%.\footnote{\url{https://jwst-docs.stsci.edu/jwst-calibration-status/nirspec-calibration-status/nirspec-ifu-calibration-status}}. 

To produce spectrally integrated emission maps, the observed data cubes in the vicinity of each emission band were first continuum-subtracted (pixel-by-pixel), using polynomial fits to the spectral baseline regions surrounding the detected emission lines.  For H$_2$O, a 7th-order polynomial was used (fitted between 2.55--2.90~$\mu$m); for CO$_2$, 1st order was used (fitted between 4.188--4.500~$\mu$m, with the CO$_2$ $\nu_3$ band masked between 4.20--4.44~$\mu$m); for CO, 4th order was used (fitted between 4.50--4.85~$\mu$m). Spectral line emission was excluded from the continuum fits using an iterative sigma-clipping algorithm with a rejection threshold 2.8 times the residual RMS.  The detected line emission was then spectrally integrated to produce the maps in Fig. \ref{fig:maps}. Although the signal-to-noise ratio (SNR) per pixel was too low to allow useful mapping of the minor isotopologues (HDO and $^{13}$CO), significant emission from these species is detected in larger extraction apertures (see Fig. \ref{fig:fits}).

\subsection*{Astrometry}

In order to guarantee the acquisition of 3I/ATLAS within the small ($3''\times3''$) NIRSpec IFU field of view, while allowing for the $0.2''$ dither offsets and $0.1''$ telescope blind pointing accuracy, we had to ensure an ephemeris accuracy better than $\pm 1.2''$ (at $3\sigma$ confidence). To reach such a high accuracy, we performed astrometric measurements of 3I/ATLAS in the weeks leading up to our JWST observations using images contributed by observers at various professional telescope facilities, located at sites with good seeing conditions (see Acknowledgements for further details). These included the Canada-France-Hawai'i Telescope, Lowell Discovery Telescope, Apache Point Observatory, Southern Astrophysical Research telescope, the Las Cumbres Observatory, and the University of Hawai'i 2.2~m telescope. Good seeing was essential to resolve the inner coma and the pseudo-nucleus with the greatest possible precision, ensuring a positional measurement as close as possible to the actual location of the cometary nucleus.

However, good resolution alone is not sufficient to derive correct and unbiased astrometry of an active comet. When asymmetric outgassing is present, the inner coma is often distorted in an antisolar direction, and most common astrometric centroiding procedures tend to follow the tailward extension of the coma, resulting in measurements that are biased away from the nucleus and towards the tail. This issue was mitigated using the comet-specific, zero aperture extrapolation method introduced by \citet{2004DPS....36.3416T} and subsequently validated by \citet{2021Icar..35814276F}. We measured various datasets obtained in the days and weeks before scheduling the JWST observations using this technique and reported all the corresponding astrometry to the Minor Planet Center, ensuring it would be available by the time of scheduling and properly included in the solution posted on the JPL Horizons database.

\subsection*{Spectral Modeling}

Gas production rates ($Q$) and rotational temperatures ($T_{rot}$), were derived as a function of distance from the nucleus for H$_2$O, CO$_2$, and CO, using optimal estimation routines as part of the Planetary Spectrum Generator (PSG; \citealt{villanueva18}, based on synthetic fluorescence models described by \citealt{Villanueva2025}). To take advantage of the spatial information in our dataset, our modeling strategy followed a similar ``$Q$-curve'' formalism as described by \cite{woo25,cor25}. Pixels close to the nucleus are affected by PSF-related flux losses and (for stronger lines), opacity effects, which are difficult to accurately model. We therefore derived gas production rates as a function of $\rho$ within successive (independent) annuli, centered on the brightest (pseudo-nucleus) pixel. Assuming uniform, isotropic gas production with a constant outflow velocity ($v_{out}$), the $Q$ value converges to an asymptote, or terminal $Q$ value ($Q_t$), at a characteristic $\rho$ value, at which point the unmodeled flux losses are negligible and $Q_t$ then represents the total molecular production rate within the field of view. To avoid contamination of these results by noisier pixels at the IFU edges, the $Q$ curves were computed up to $\rho=1.8''$, corresponding to the $3.6''$-diameter circular integration apertures used in Figure \ref{fig:fits}. The gas outflow velocity was set to 0.310~km\,s$^{-1}$ for H$_2$O and CO$_2$, and 0.345~km\,s$^{-1}$ for CO, based on contemporaneous observations of the spectrally resolved HCN and CO line shapes, using the ALMA Atacama Compact Array (see ALMA Observations section). Molecular photolysis rates appropriate for the active Sun were incorporated from \citet{hue15}. The spectral resolution for each NIRSpec grating (as a function of wavelength) was taken from the dispersion curves available at https://jwst-docs.stsci.edu.

Parameter uncertainties were derived from the diagonal elements of the fit covariance matrix, scaled according to the square root of the reduced chi-square ($\chi_R^2$) to ensure the baseline noise level was realistically accounted for. For all species apart from CO$_2$ the spectral baseline shape (with contributions due to scattered sunlight and thermal emission from the coma dust and nucleus, combined in some cases with instrumental artifacts), was determined by fitting a polynomial function simultaneously with the gas emission lines. This way, statistical uncertainties in the baseline shape are propagated into the final uncertainties on each $Q$ value. For CO$_2$, the inaccurate model fit to some of the higher-$J$ CO$_2$ lines (possibly due to the presence of multiple $T_{rot}$ components or non-LTE effects), results in the baseline fit becoming skewed in the direction of the fit residuals. Therefore, prior to performing the CO$_2$ $Q$-curve analysis, it was necessary to subtract a fitted linear baseline from these data pixel-by-pixel (with the CO$_2$ + $^{13}$CO$_2$ line-containing region masked between 4.20--4.44~$\mu$m). Due to the high CO$_2$ line-to-continuum ratio ($\sim200$), uncertainties in the baseline placement for this species have negligible impact on its retrieved production rate.

The H$_2$O band at 2.7~$\mu$m (in the G235H setting) contains multiple, high-SNR, spectrally resolved rovibrational lines of the ortho (o) and para (p) H$_2$O spin states (with H-atom spins parallel \emph{vs.} antiparallel, respectively), so this band was used to determine $T_{rot}$(H$_2$O) and the H$_2$O ortho-to-para ratio (OPR) --- key parameters for reliably measuring $Q$(H$_2$O) from the weaker, more blended 4.5--5.1~$\mu$m H$_2$O hot bands that were observed simultaneously with HDO in the G395H setting. We modeled o- and p-H$_2$O simultaneously, but as separate species, using the NASA PSG, with $0.2''$-wide annuli extracted with increasing nucleocentric distance, allowing $Q$(o-H$_2$O), $Q$(p-H$_2$O), and $T_{rot}$ to vary as free parameters. An example of the spectral fit quality is shown in Extended Data Fig. \ref{fig:opr}.

The terminal (asymptotic) p-H$_2$O production rate at $\rho=1.8''$ is $Q_t=(4.28\pm0.07)\times10^{26}$ s$^{-1}$, while for o-H$_2$O, $Q_t=(1.19\pm0.01)\times10^{27}$ s$^{-1}$, leading to OPR = $2.77\pm0.05$ (close to the statistical equilibrium value of 3), with a total (terminal) water production rate $Q_t$(H$_2$O) = $(1.61\pm0.01)\times10^{27}$ s$^{-1}$, summed over the ortho and para contributions. Due to the large o- and p-H$_2$O emission line strengths compared with the continuum around 2.7~$\mu$m, these results are not significantly dependent on the choice of baseline polynomial order.

The H$_2$O hot bands around 5~$\mu$m contained insufficient unblended lines for a reliable $T_{rot}$ and OPR retrieval, but since they were observed simultaneously with HDO, the hot band lines were used to derive $Q_t$(H$_2$O) for direct comparison with $Q_t$(HDO). For the H$_2$O hot band (and HDO), the OPR was fixed at 2.77 and the rotational temperature for each annulus ($T_{rot}(\rho)$) was held fixed according to the $T_{rot}(\rho)$ values previously derived from the H$_2$O  2.7~$\mu$m band. Many of the H$_2$O hot-band lines in our data are blended with lines from CO, CN and OCS, so to derive $Q$(H$_2$O), we focused on the 4.5--4.7~$\mu$m region, which contains three of the stronger, unblended H$_2$O lines (see Fig. \ref{fig:fits}a). Prior to fitting the H$_2$O spectrum in each annulus, we subtracted the interloping CO lines, using a set of independent Gaussian fits. This facilitates identification of the spectral baseline and helps remove contamination from any imperfectly-fit CO lines in the spectral model; the spectral regions within the boundaries of each subtracted CO line (see Fig. \ref{fig:fits}a) were further masked to prevent any remaining CO residuals from contributing to the fit. The resulting $Q_{\rho}$(H$_2$O) curve is shown in Extended Data Fig. \ref{fig:qcurves}. Taking the average of the outer 5 annuli gives $Q_t({\rm H_2O})=(1.69 \pm 0.05)\times10^{27}$ s$^{-1}$, which is consistent with the value obtained from the G235H setting on the previous day.

Four individual rovibrational lines of HDO were clearly detected, as part of the $\nu_1$ band between $\lambda=3.55$--3.85~$\mu$m (Fig. \ref{fig:fits}b). No evidence for other significant emission lines, including OH prompt emission, could be identified in this spectral region. A possible weak line of H$_2$CO could be present at 3.689~$\mu$m, just to the red of the 3.692~$\mu$m HDO line, but H$_2$CO contamination of the HDO signal is expected to be negligible, considering that the expected (stronger) H$_2$CO lines between 3.58--3.61~$\mu$m are not readily apparent in our data. Due to a relatively complex continuum shape in this region, a 7th-order polynomial was required to obtain a good fit to the baseline. The HDO rotational temperature within a $d=3''$ circular aperture ($16\pm3$~K) was found to be consistent with that of H$_2$O at 2.7~$\mu$m, which justifies the application of the $T_{rot}(\rho)$(H$_2$O; 2.7~$\mu$m) values during the HDO fitting. Due to the large number of HDO lines and the lack of significant blending by other species, our derived $Q$(HDO) values are accurate to $15\sigma$ confidence. Taking the average of the outer 5 annuli of the $Q$-curve (Extended Data Fig. \ref{fig:qcurves}) results in $Q_t({\rm HDO})=(3.20 \pm 0.19)\times10^{25}$ s$^{-1}$. This corresponds to $Q_t({\rm HDO})/Q_t({\rm H_2O}) = (1.90 \pm 0.13)\times10^{-2}$, or D/H = $(0.95\pm0.06)$\% (taking into account the number of hydrogen atoms in H$_2$O).

To determine $Q_{\rho}$(CO$_2$) and $T_{rot}(\rho)$(CO$_2$), the continuum-subtracted NIRSpec data cube was simultaneously fit for CO$_2$ and $^{13}$CO$_2$ components in the range 4.15--4.50~$\mu$m. These values were then held fixed to derive $Q_{\rho}$($^{13}$CO$_2$) using the non-continuum-subtracted cube (confined within the spectral range 4.34--4.50~$\mu$m), but this time including a variable polynomial baseline in the PSG retrieval. Continuum definition is nontrivial in the vicinity of the $^{13}$CO$_2$ band due to the presence of a relatively broad, unidentified spectral feature on the long-wavelength side of the $^{13}$CO$_2$ P-branch (around 4.42~$\mu$m), combined with the close proximity of the main CO$_2$ band on the short-wavelength side of the $Q$-branch, for which the high-$J$ lines are not accurately modeled. We therefore investigated the effects of continuum uncertainty on $Q_t$($^{13}$CO$_2$) by generating an ensemble of $Q$-curve fits, each with a different polynomial baseline order (between 1--5). Due to potential contamination of the $^{13}$CO$_2$ P-branch by the aforementioned unidentified spectral feature, we also performed a separate ensemble of fits, masking the region between $\lambda=4.380$--$4.445$~$\mu$m (to exclude the $^{13}$CO$_2$ P-branch as well as the unidentified feature). The resulting range of fitted baselines is shown in Extended Data Fig. \ref{fig:bases}a. The corresponding $Q_t$($^{13}$CO$_2$) values (and associated $^{12}$C/$^{13}$C ratios) for each of the fitted baselines are shown in Table \ref{tab:13co2}, along with the reduced $\chi^2$ values ($\chi^2_R$) for the (model minus fit) residuals. From the best-fitting $Q_t$(CO$_2$) value of $(1.76 \pm 0.01)\times10^{27}$ s$^{-1}$ combined with the range of possible $Q_t$($^{13}$CO$_2$) values, we derive $Q_t$(CO$_2$)/$Q_t$($^{13}$CO$_2$) = 141--191 (including the $1\sigma$ statistical error margins).
 
The region between $\lambda=4.5$--5.1~$\mu$m contains the strong $v=1-0$ band of CO, in addition to a blend of $^{13}$CO, H$_2$O, CN and OCS emission features. Retrieving the $^{12}$CO/$^{13}$CO ratio therefore requires careful modeling of all these features. Similar to CO$_2$, the main CO band is strong enough that baseline uncertainties have negligible impact on its retrieved production rate. We therefore initially performed a $Q$-curve analysis for CO by fitting the production rates for all species as a function of $\rho$, in addition to retrieving $T_{rot}(\rho)$(CO) (see Extended Data Figs. \ref{fig:qcurves} and \ref{fig:trots}), with a 5th-order polynomial baseline included in the PSG fit. The resulting terminal production rate was $Q_t({\rm CO}) = (3.94 \pm 0.01)\times10^{27}$ s$^{-1}$. The fit to the CO $v=1-0$ band is very good apart from in the highest-$J$ CO lines, some of which are blended with lines of $^{13}$CO (Fig. \ref{fig:fits}e). Therefore, to retrieve $Q_{\rho}$($^{13}$CO), we focused on the $\lambda=4.71$--5.05~$\mu$m region (which contains 15 individual lines of $^{13}$CO), with the poorly-fitting higher-$J$ CO lines masked out so that those regions did not influence the fit (see Fig. \ref{fig:fits}f). We performed the $Q$-curve analysis for $^{13}$CO by fixing the rotational temperatures according to the previously retrieved $T_{rot}(\rho)$(CO) curve, with polynomial baseline orders between 1--5; in each case, the polynomial coefficients were optimized during fitting so that their uncertainties propagated into the final error estimates on $Q_{\rho}$($^{13}$CO). The resulting range of fitted baselines is shown in Extended Data Fig. \ref{fig:bases}b. Due to the weakness of the lines, the $^{13}$CO signal became too noisy toward the edge of the field of view, so the last two annuli were omitted from the $^{13}$CO $Q$-curve analysis (Extended Data Fig. \ref{fig:qcurves}). The best-fitting $Q_t$($^{13}$CO) value and associated $^{12}$C/$^{13}$C ratio and $\chi^2_R$ for each baseline order is given in Table \ref{tab:13co}, corresponding to a range of $^{12}$CO/$^{13}$CO values between 123--172 (including $1\sigma$ statistical error margins).

Taking the average of the last 5 points in the $Q$-curves for CN and OCS, from our fits to the $\lambda=4.5$--5.1~$\mu$m region, the resulting $Q_t$(CN) and $Q_t$(OCS) values for each baseline order are given in Table \ref{tab:13co}.  Averaged over the set of 5 different baseline orders, we derive $Q_t({\rm CN})=(3.10 \pm 0.6)\times10^{24}$ s$^{-1}$ and $Q_t({\rm OCS})=(4.70 \pm 0.11)\times10^{24}$ s$^{-1}$, where the errors are the average statistical uncertainties combined in quadrature with the standard deviations of the respective $Q_t$ values (across the 5 different baseline orders). $Q_t({\rm CN})$ should be considered a strict lower limit because CN is produced in the comae of Solar System comets with a radial scale length $\sim3\times10^4$~km \citep{fra05}, so the majority of its production is expected to be outside of the NIRSpec field of view.

The measured OCS/CO$_2$ production rate ratio is $(0.27\pm0.03)$\%. This is significantly less than the values of 0.6--0.9\% found in comet 67P/Churyumov-Gerasimenko \citep{boc16}, and may indicate a relatively low elemental S/O value for the 3I/ATLAS originating system.

\subsection*{ALMA Observations}

Submillimeter interferometric observations of comet 3I/ATLAS were obtained with the Atacama Compact Array (ACA) sub-component of the Atacama Large Millimeter/submillimeter Array (ALMA) as part of ALMA project 2025.A.00004.S. The observations were conduced in nominal Band 7 weather conditions on 2025-Dec-22 between UT 07:32--08:36, using $11\times7$~m antennas, with a median zenith precipitable water vapour column of 0.33~mm. The Band 7 receiver was configured to simultaneously observe the CO $J=3-2$ line at 345795.990~MHz at a spectral resolution of 122~kHz and the HCN $J=4-3$ line at 354505.4773~MHz at a spectral resolution of 61~kHz. The position and radial velocity of the object were tracked using JPL Horizons solution \#44.  The interferometric data were flagged and calibrated in the Common Astronomy Software Applications (CASA) package \citep{casa22} using standard scripts supplied by the Joint ALMA Observatory. The interferometric point-spread function was then deconvolved from the spectral image data using the CASA {\tt tclean} Hogb{\"o}m algorithm with natural weighting, using a mask size equal to the primary beam FWHM and stopping at a threshold of twice the RMS noise. The resulting (2D Gaussian) beam FWHM was $5.08''\times3.37''$ for CO and $4.91''\times3.28''$ for HCN.

The molecular spectra were extracted from the CO and HCN integrated emission peak position, then subjected to modeling using the SUBLIME radiative transfer code \citep{cor22}, in order to derive the gas outflow velocities ($v_{out}$). State-to-state collisional rates for H$_2$O-CO and H$_2$O-CN were taken from \citet{fau20} and \citet{zol25}, respectively, with photolysis rates for the active Sun from \citet{hue15} and solar pumping rates from \citet{cor22,cor23}. For the HCN $J=4-3$ line, hyperfine components were included as described by \citet{cor24}. Our calculations assumed $Q_t({\rm H_2O})=1.7\times10^{27}$ s$^{-1}$ and $T_{kin}=40$~K (based on $T_{rot}=40.3 \pm 0.3$~K retrieved from our JWST CO data within a $d=3''$ circular aperture).  Ideally, collisions with CO and CO$_2$ should also be considered in the SUBLIME molecular excitation calculation, but in reality, the retrieved outflow velocities are not affected by the assumed collision rates. Because the gas coma of 3I/ATLAS is larger than the maximum recoverable scale for the ACA ($\sim19''$), interferometric flux losses were modeled by applying a constant flux loss factor of 0.75 for the central beam, which was derived by comparing the peak flux of our model CO image before and after processing it using the CASA {\tt simobserve} task. 

Assuming a spherically-symmetric coma, the best-fitting SUBLIME models gave $v_{out}({\rm CO})=0.345\pm0.010$~km\,s$^{-1}$; $Q({\rm CO})=(2.64 \pm 0.01)\times10^{27}$ s$^{-1}$, and, $v_{out}({\rm HCN})=0.276\pm0.015$~km\,s$^{-1}$; $Q({\rm HCN})=(6.75 \pm 0.54)\times10^{24}$ s$^{-1}$ (see Extended Data Fig. \ref{fig:almadata}). These gas outflow velocities are smaller than the values $\sim0.5$~km\,s$^{-1}$ typically found in comets at similar heliocentric distances, which can be explained as a result of outgassing driven by a species with a higher molecular weight than H$_2$O (Biver et al. 2026, submitted). This is consistent with the high CO$_2$/H$_2$O and CO/H$_2$O abundances (greater than unity) measured in 3I/ATLAS by JWST and ALMA, implying that CO and CO$_2$ sublimation are the main drivers of outgassing in this comet. Since the region of the coma where the gases are accelerated (close to the nucleus) is typically in collisional equilibrium \citep{zak23}, different molecules are expected to share a common outflow velocity. Differences could occur as a result of angular variations in the coma composition, or a different balance of production rates from nucleus \emph{vs.} icy grain sublimation for different species. Therefore, to model our JWST CO$_2$ and H$_2$O data (for which spectrally resolved line profiles were not available), we take the average of the CO and HCN outflow velocities: $v_{out}=0.310$~km\,s$^{-1}$, while the observed value of $v_{out}=0.345$~km\,s$^{-1}$ was used for CO.  Ultimately, the use of the $Q$-curve formalism for deriving isotopic ratios helps minimize optical depth effects, such that uncertainties in $v_{out}$ have a linear effect on our presented absolute molecular production rates. Consequently, $v_{out}$ uncertainties should have negligible impact on the retrieved isotopic ratios.

%If your article has accompanying supplementary file/s please state so here. Please refer to Journal-level guidance for any specific requirements.

\bmhead{Acknowledgements}

This work is based on observations made with the NASA/ESA/CSA James Webb Space Telescope. The data were obtained from the Mikulski Archive for Space Telescopes at the Space Telescope Science Institute, which is operated by the Association of Universities for Research in Astronomy, Inc., under NASA contract NAS 5-03127 for JWST. We gratefully acknowledge the assistance of optical observers who submitted astrometric observations of 3I/ATLAS in the weeks leading up to our observations, to help refine the ephemeris position. In particular, we thank J. Chatelain, E. Gomez, S. Greenstreet, W. Hoogendam, C. Holt, H. W. Lin, T. Lister, T. Santana-Ros, L. Salazar Manzano, D. Seligman, Q. Ye, Q. Zhang, M. Frissell and D. Singh. Supporting astrometric observations were obtained by the Comet Chasers school outreach program (https://www.cometchasers.org/ ), led by Helen Usher, which is funded by the UK Science and Technology Facilities Council (via the DeepSpace2DeepImpact Project), the Open University and Cardiff University. It accesses the LCOGT telescopes through the Schools Observatory/Faulkes Telescope Project (TSO2025A-00 DFET-The Schools’ Observatory), which is partly funded by the Dill Faulkes Educational Trust, and through the LCO Global Sky Partners Programme (LCOEPO2023B-013). Observers included individuals and representatives from the following schools and clubs: E. Maciulis, A. Bankole, J. Bower, O. Roberts, participants on the British Astronomical Associations’ Work Experience project 2025 from The Coopers Company \& Coborn School, Upminster, UK;  St Marys Catholic Primary School, Bridgend, UK; J. M. Perez Redondo \& Students: A. Matea, L. Guillamet, A. Montoy, and A. Martin from Institut d’Alcarràs, Catalonia, Spain; Louis Cruis Astronomy Club, Brazil; Jelkovec High School, Zagreb, Croatia, and C. Wells at a British Astronomical Association event. ALMA is a partnership of ESO, NSF (USA), NINS (Japan), NRC (Canada), NSC and ASIAA (Taiwan) and KASI (Republic of Korea), in cooperation with the Republic of Chile. The JAO is operated by ESO, AUI/NRAO and NAOJ.  The NRAO is a facility of the National Science Foundation operated under cooperative agreement by Associated Universities, Inc.. M.A.C., N.X.R., S.B.C. and S.N.M. were supported by the NASA Planetary Science Division Internal Scientist Funding Program through the Fundamental Laboratory Research work package (FLaRe). C.O.C. acknowledges support from the DIRAC Institute in the Department of Astronomy at the University of Washington.  D.F. conducted this research at the Jet Propulsion Laboratory, California Institute of Technology, under a contract with the National Aeronautics and Space Administration (80NM0018D0004). M.E.S. acknowledges support in part from UK Science and Technology Facilities Council (STFC) grant ST/X001253/1.  This research has made use of NASA’s Astrophysics Data System Bibliographic Services. This research has made use of data and/or services provided by the International Astronomical Union's Minor Planet Center.

%\section*{Declarations}

%If any of the sections are not relevant to your manuscript, please include the heading and write `Not applicable' for that section. 

\begin{itemize}
\item {\bf Data Availability:}
All JWST data are available through the Mikulski Archive for Space Telescopes at the Space Telescope Science Institute under proposal ID \#5094 (https://doi.org/10.17909/1jvn-1z72). The JWST data are under a three month embargo from the date of acquisition. 
%This work makes use of ALMA data set ADS/JAO.ALMA\#2025.A.00004.S, which is available for download from the ALMA Science Archive (http://almascience.nrao.edu/aq/).
\item {\bf Code Availability:}
The JWST Calibration Pipeline software \citep{bushouse25} is available from https://github.com/spacetelescope/jwst.

The Planetary Spectrum Generator \citep{villanueva18}, used for modeling cometary infrared emission lines  is available at https://psg.gsfc.nasa.gov/.

The {\tt jwstComet} software \citep{Roth2026}, used for extracting spectra from JWST data cubes and modeling them with PSG, is available from https://github.com/apertureSynthesis/jwstComet.

The 1D version of the SUBLIME radiative transfer code \citep{cor22}, used for modeling the ALMA data, is available from https://github.com/mcordiner/sublime-d1dc.  The {\tt sublimed1dFit} code for generating optimised spectral line models using SUBLIME is available from https://github.com/mcordiner/sublimed1dFit.

\item {\bf Author Contributions:} M. Cordiner calibrated the JWST and ALMA data, performed the NIRSpec G395H spectral extraction and modeling, and led the manuscript writing. N. Roth and G. Villanueva contributed to the data analysis methodology, performed the G235H (2.7~$\mu$m) H$_2$O band modeling, and independently checked the isotopic ratio retrievals. M. Micheli performed astrometric measurements. D. Farnocchia calculated the 3I/ATLAS orbit and ephemeris. S. Charnley helped interpret the isotopic ratios. All authors helped with the project design, data acquisition, interpretation of results, and editing of the manuscript.

\item {\bf Competing Interests:} The authors declare no competing interests.

\end{itemize}

\clearpage
\setcounter{figure}{0}
\renewcommand{\figurename}{Extended Data Figure}
\setcounter{table}{0}
\captionsetup[table]{name=Extended Data Table}

\section*{Extended Data}

\begin{figure*}
\begin{center}
\includegraphics[height=6.5cm]{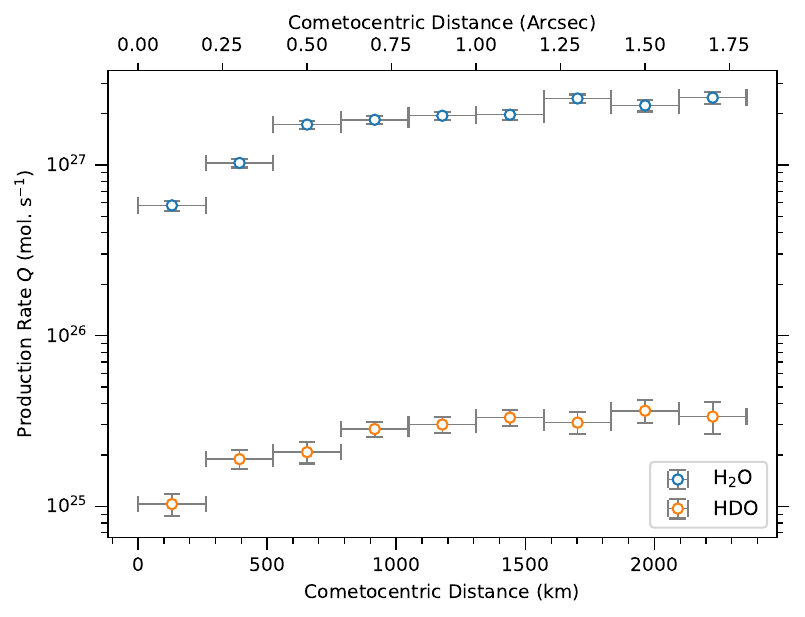}
\includegraphics[height=6.5cm]{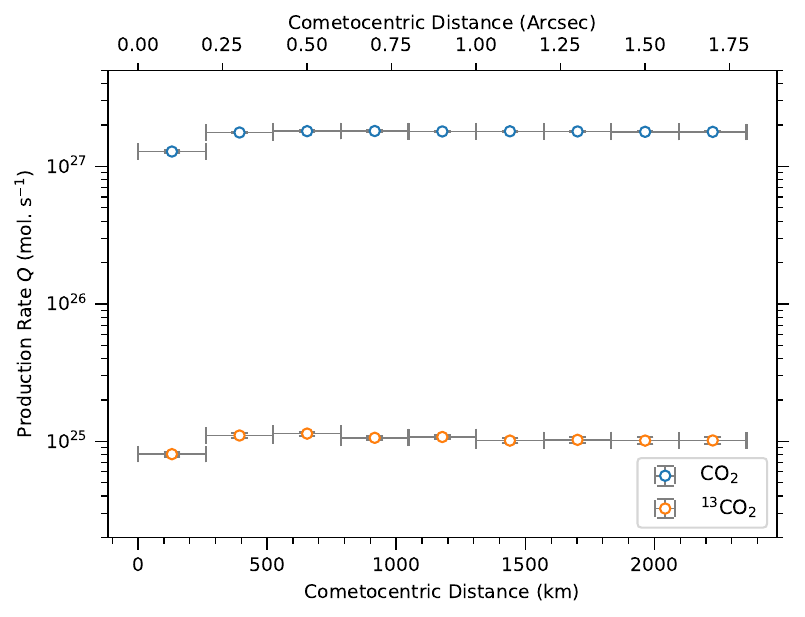}
\includegraphics[height=6.5cm]{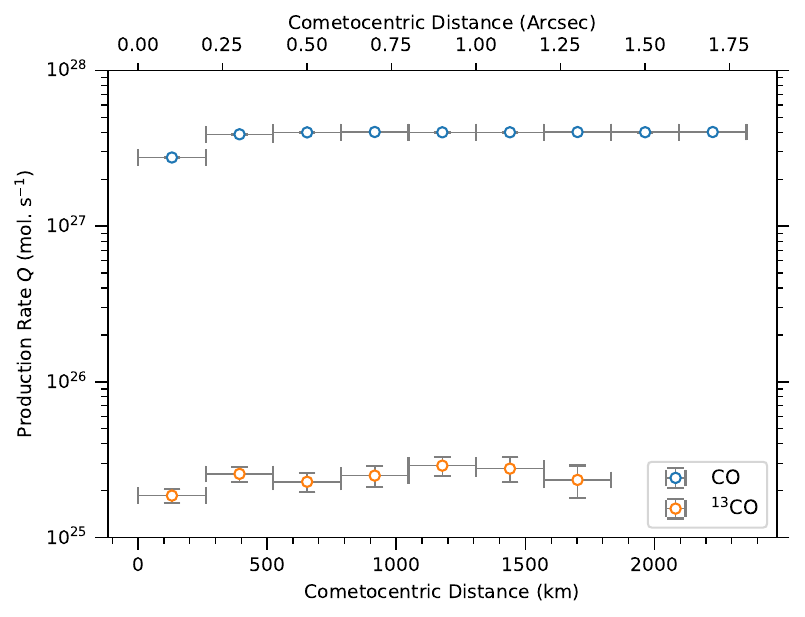}
\end{center}
\caption{Best fitting production rates ($Q$) for H$_2$O, CO$_2$, CO and their minor isotopologues, as a function of sky-projected distance ($\rho$). Spectra were extracted and modeled within successive $0.2''$-thick annuli centered on the coma flux peak to generate $Q_{\rho}$. For $^{13}$CO$_2$ and $^{13}$CO, the models included a 5th-order polynomial baseline fit. Vertical error bars indicate $1\sigma$ statistical uncertainties, while the horizontal error bars indicate the radial extent of each spatial region.}
\label{fig:qcurves}
\end{figure*}

\begin{figure*}
\begin{center}
\includegraphics[width = 0.49\textwidth]{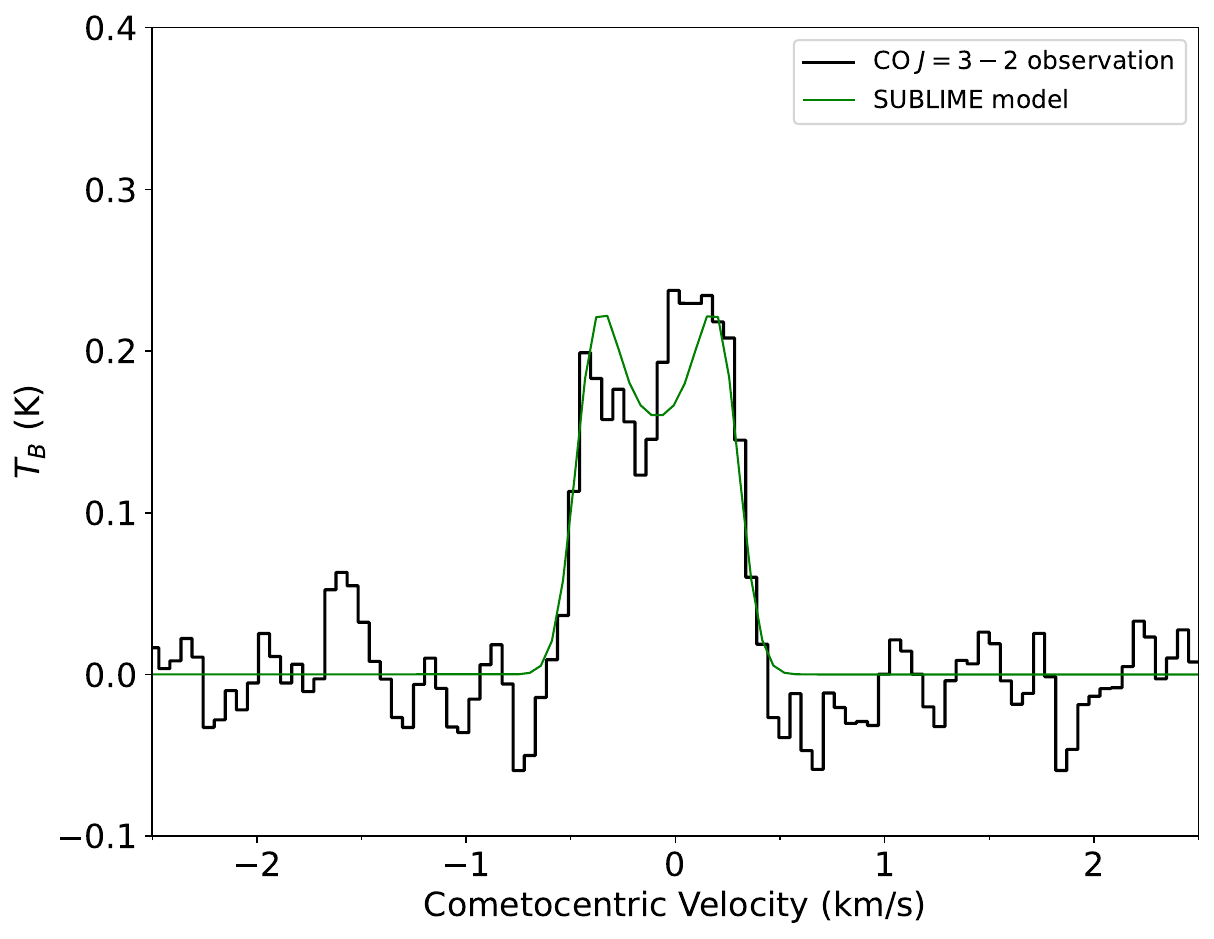}
\includegraphics[width = 0.49\textwidth]{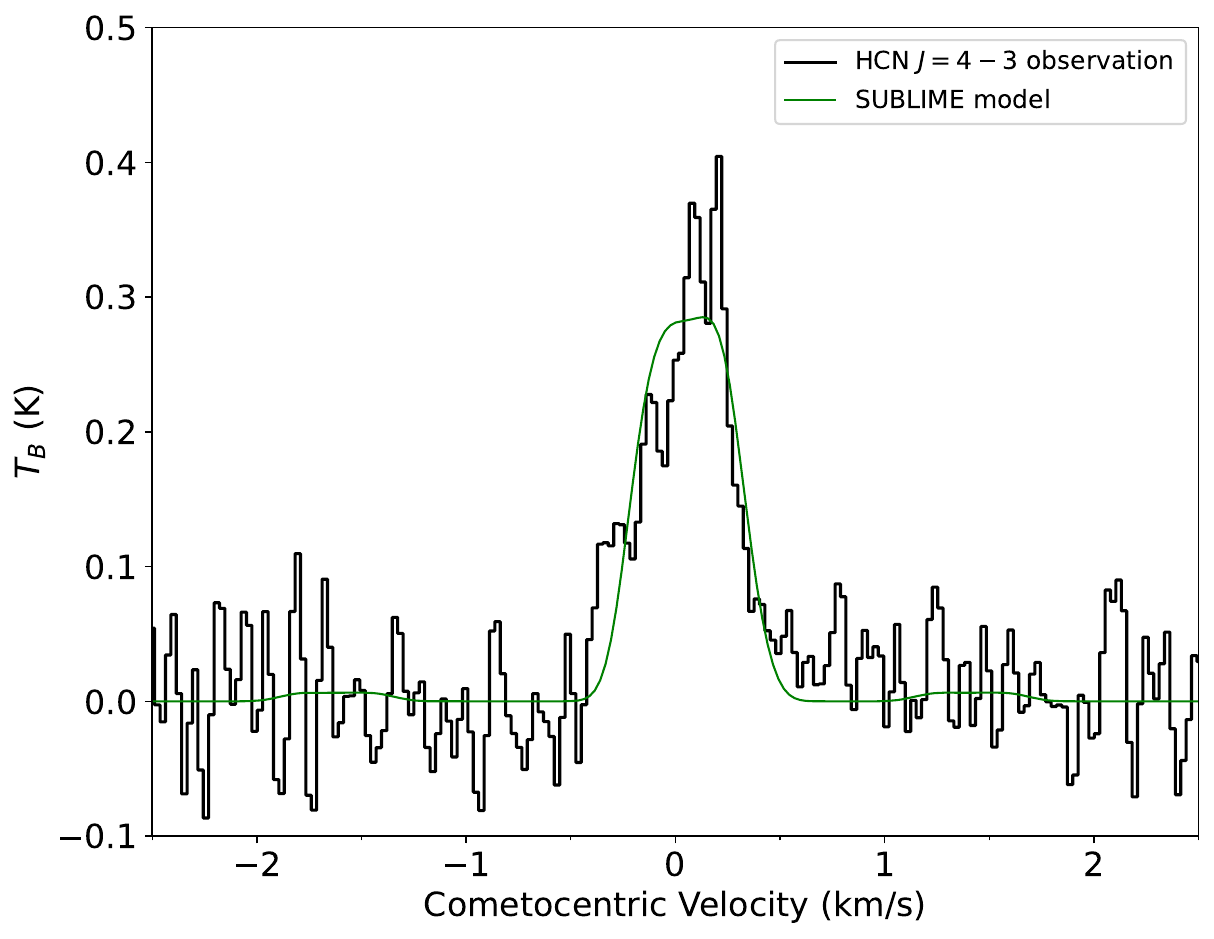}
\end{center}
\caption{High-resolution submillimeter-wave spectra of the CO (left) and HCN (right) line emission from 3I/ATLAS, observed using the ALMA ACA on 2025-Dec-22. Spectra were extracted at the common emission peak for both species, and are plotted on a velocity scale with respect to the rest frequencies of the CO $J=3-2$ and HCN $J=4-3$ lines, respectively. Best fitting spectral models are overlaid.}
\label{fig:almadata}
\end{figure*}

\begin{figure*}
\begin{center}
\includegraphics[width=0.6\textwidth]{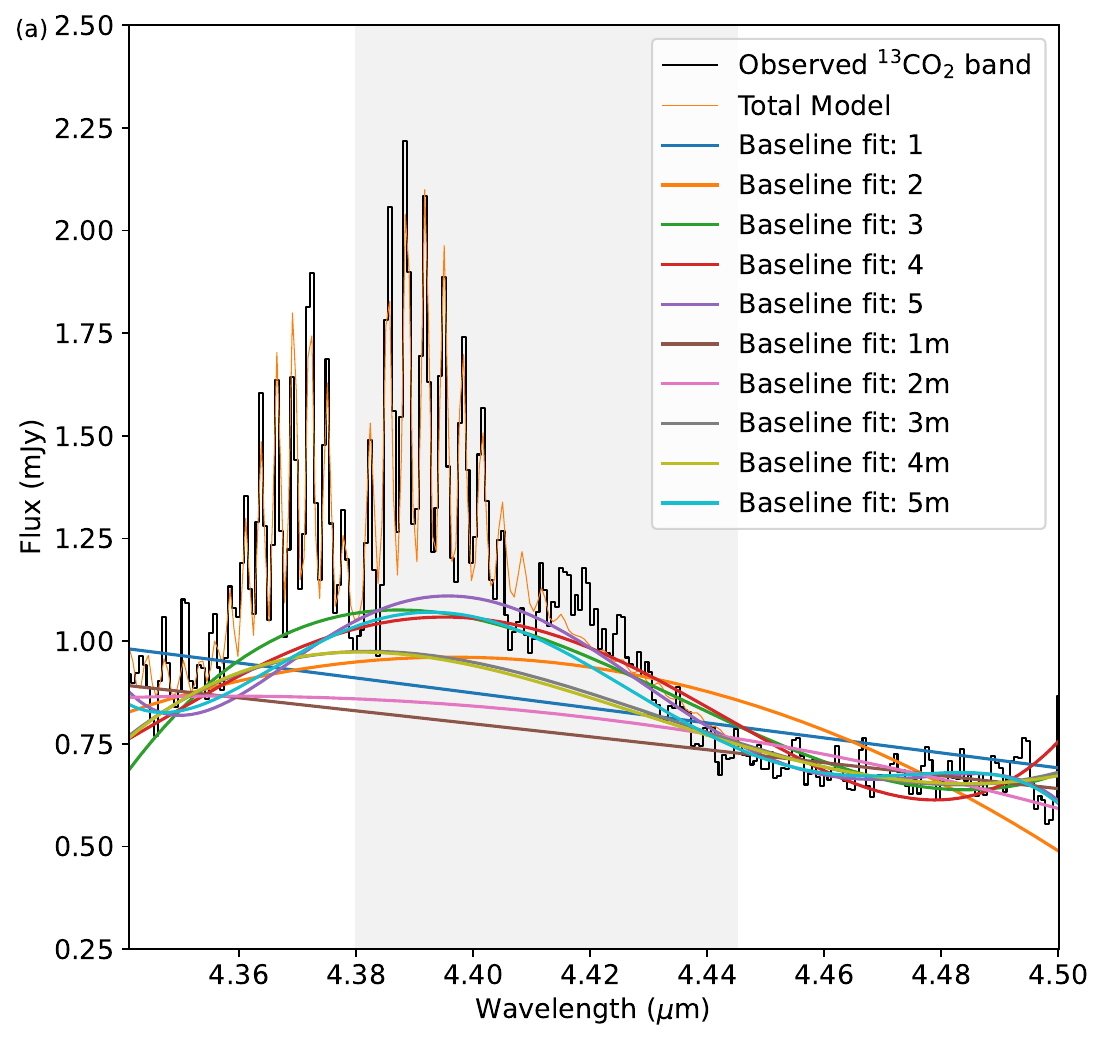}
\includegraphics[width=0.6\textwidth]{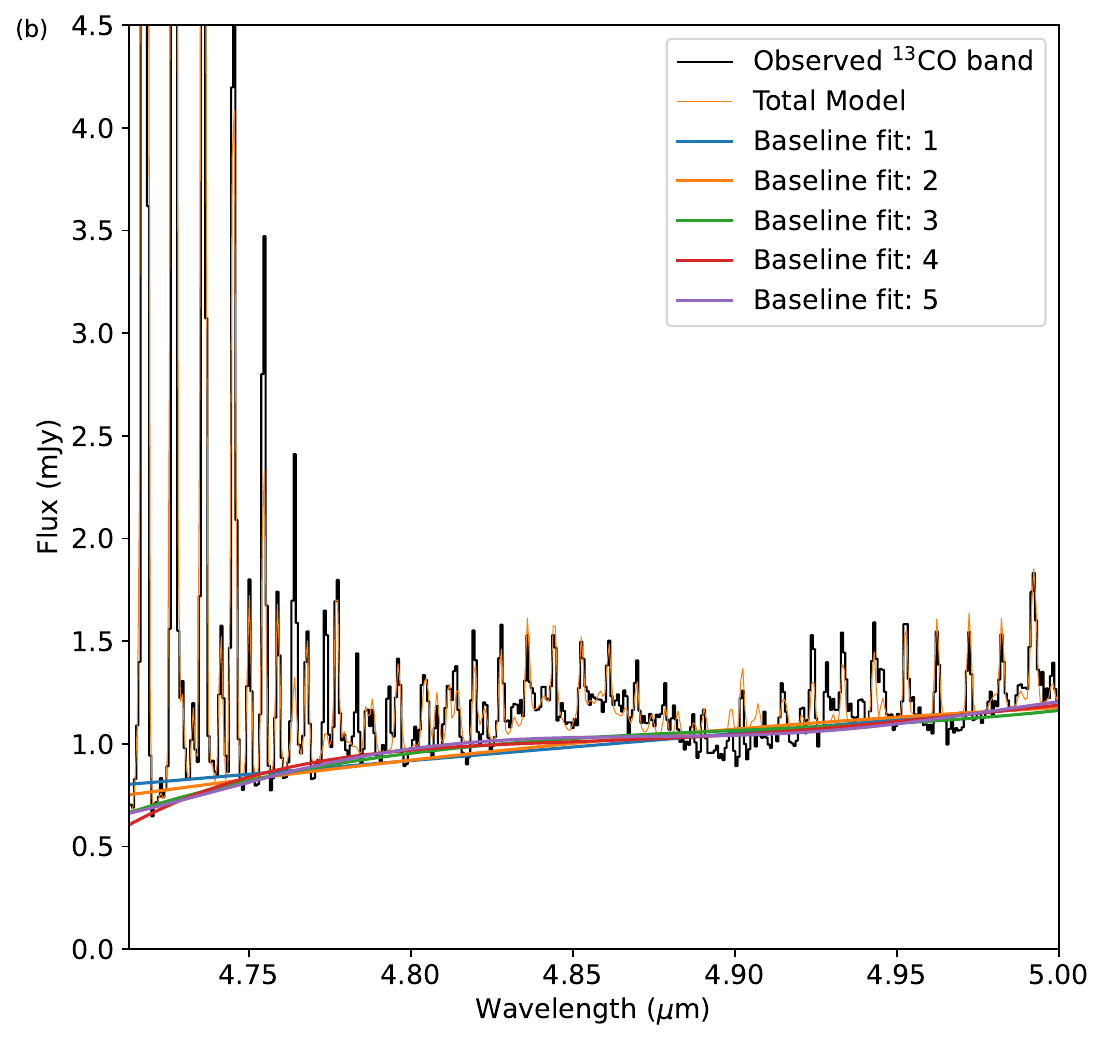}
\end{center}
\caption{3I/ATLAS NIRSpec spectrum in the vicinity of (a) the $^{13}$CO$_2$ $\nu_3$ band and (b) the $^{13}$CO $v=1-0$ band, integrated within a $d=3.6''$ circular aperture. The ensemble of best-fitting polynomial baselines considered as part of our spectral modeling is overlaid, corresponding to the $Q_t$($^{13}$CO$_2$) and $Q_t$($^{13}$CO) values shown in Tables \ref{tab:13co2} and \ref{tab:13co}, respectively.  For those fits labeled `m' in panel (a), the potentially contaminated region between $\lambda=4.380$--$4.445$~$\mu$m (highlighted in grey) was excluded (masked) during fitting. For both panels, the ``total model'' curves are for the minimum $\chi^2_R$ model (using a 5th order baseline in both cases). For $^{13}$CO, numerous emission lines from additional species are present (see Fig. \ref{fig:fits}f for details). }
\label{fig:bases}
\end{figure*}

\begin{figure*}[bh!]
\begin{center}
\includegraphics[width=0.8\textwidth]{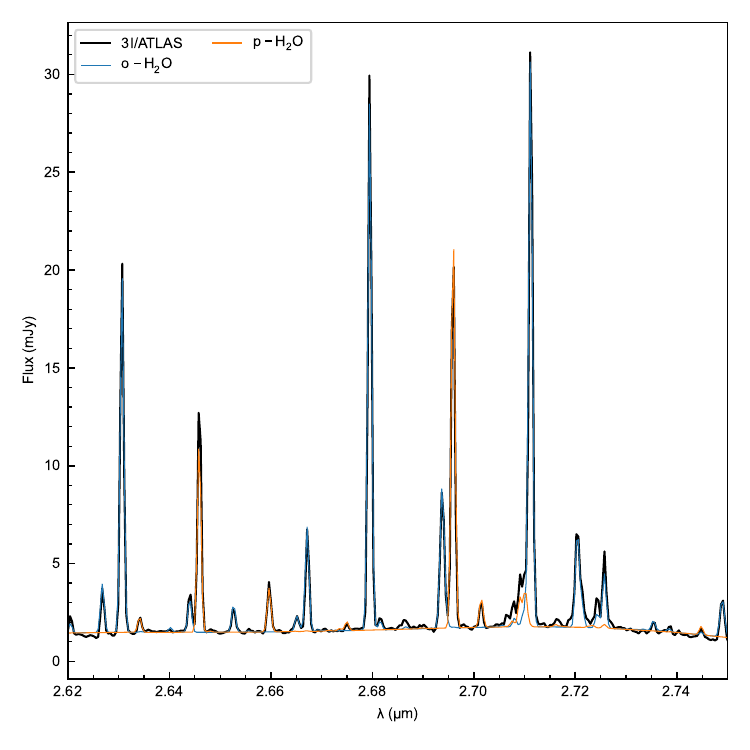}
\end{center}
\caption{NIRSpec H$_2$O 2.7~$\mu$m spectrum of 3I/ATLAS observed 2025-Dec-22, extracted inside a $d=3.0''$ circular aperture, with independent ortho- and para-H$_2$O spectral fits overlaid.}
\label{fig:opr}
\end{figure*}

\begin{figure*}
\begin{center}
\includegraphics[width=0.6\textwidth]{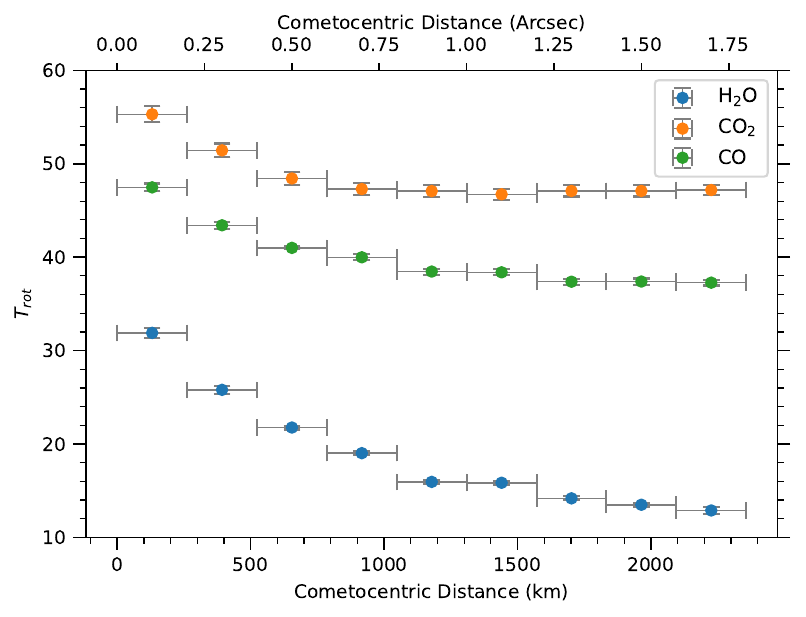}
\end{center}
\caption{Best fitting molecular rotational temperatures ($T_{rot}$) as a function of $\rho$. Vertical error bars indicate $1\sigma$ statistical uncertainties, while the horizontal error bars indicate the radial extent of each spatial region.}
\label{fig:trots}
\end{figure*}

\clearpage

\begin{table}
\caption{\label{tab:13co2} $^{13}$CO$_2$ terminal gas production rates and $^{12}$CO$_2$/$^{13}$CO$_2$ ratios for different baseline fit orders}
\centering
\begin{tabularx}{8cm}{XXcX}
\hline\hline
Order & $Q_t$($^{13}$CO$_2$) & $\chi^2_R$& $^{12}$C/$^{13}$C\\
 & ($10^{25}$~s$^{-1}$)  &                &\\
\hline
1&    $1.22 \pm 0.03$ & 9.956     &    $144\pm3$      \\      
2&    $1.10 \pm 0.02$ & 6.847     &    $161\pm3$      \\
3&    $1.15 \pm 0.02$ & 4.228     &    $154\pm3$      \\
4&    $1.05 \pm 0.02$ & 3.853      &    $168\pm4$      \\
5&    $1.02 \pm 0.02$ & 3.022      &    $172\pm4$      \\[2mm]
1(m)& $1.11 \pm 0.02$ &13.70   &    $158\pm3$      \\
2(m)& $1.10 \pm 0.02$ &10.74   &    $161\pm3$      \\
3(m)& $0.97 \pm 0.02$ &6.759   &    $182\pm4$      \\
4(m)& $0.95 \pm 0.02$ &7.218   &    $186\pm4$      \\
5(m)& $0.94 \pm 0.02$ &3.536   &    $187\pm4$      \\
\hline
\end{tabularx}
\parbox{8cm}{Note --- (m) indicates the fit was performed with the $^{13}$CO$_2$ P-branch masked between $\lambda=4.380$--$4.445$~$\mu$m to avoid baseline contamination.}
\end{table}

\begin{table}
\caption{\label{tab:13co} $^{12}$CO/$^{13}$CO ratios and $^{13}$CO, CN and OCS terminal gas production rates for different baseline fit orders}
\centering
\begin{tabular}{lccccc}
\hline\hline
Order & $Q_t$($^{13}$CO) & $Q_t$(CN) & $Q_t$(OCS) & $\chi^2_R$& $^{12}$C/$^{13}$C\\
 & ($10^{25}$~s$^{-1}$)  &  ($10^{24}$~s$^{-1}$)  & ($10^{24}$~s$^{-1}$)  &              &\\
\hline
1& $2.90 \pm 0.22$ & $3.18 \pm 0.07$ &$5.36 \pm 0.26$ &1.761     &    $136\pm10$      \\      
2& $2.99 \pm 0.22$ & $3.11 \pm 0.07$ &$4.92 \pm 0.26$ &1.702     &    $133\pm10$      \\
3& $2.69 \pm 0.21$ & $3.12 \pm 0.07$ &$4.18 \pm 0.26$ &1.562     &    $147\pm12$      \\
4& $2.53 \pm 0.21$ & $3.07 \pm 0.07$ &$4.72 \pm 0.27$ &1.463      &    $156\pm13$      \\
5& $2.49 \pm 0.21$ & $3.05 \pm 0.07$ &$4.32 \pm 0.29$ &1.426      &    $158\pm14$      \\
\hline
\end{tabular}
\end{table}

%\begin{appendices}

%\section{Section title of first appendix}\label{secA1}




%\end{appendices}


\clearpage

\bibliography{refs}% common bib file

@ARTICLE{sla24,
       author = {{Slavicinska}, Katerina and {van Dishoeck}, Ewine F. and {Tychoniec}, {\L}ukasz and {Nazari}, Pooneh and {Rubinstein}, Adam E. and {Gutermuth}, Robert and {Tyagi}, Himanshu and {Chen}, Yuan and {Brunken}, Nashanty G.~C. and {Rocha}, Will R.~M. and {Manoj}, P. and {Narang}, Mayank and {Megeath}, S. Thomas and {Yang}, Yao-Lun and {Looney}, Leslie W. and {Tobin}, John J. and {Beuther}, Henrik and {Bourke}, Tyler L. and {Linnartz}, Harold and {Federman}, Samuel and {Watson}, Dan M. and {Linz}, Hendrik},
        title = "{JWST detections of amorphous and crystalline HDO ice toward massive protostars}",
      journal = {\aap},
         year = 2024,
        month = aug,
       volume = {688},
          eid = {A29},
        pages = {A29},
          doi = {10.1051/0004-6361/202449785},
archivePrefix = {arXiv},
       eprint = {2404.15399},
 primaryClass = {astro-ph.SR},
       adsurl = {https://ui.adsabs.harvard.edu/abs/2024A&A...688A..29S}
}

@ARTICLE{ram14,
       author = {{Ramstedt}, S. and {Olofsson}, H.},
        title = "{The $^{12}$CO/$^{13}$CO ratio in AGB stars of different chemical type. Connection to the $^{12}$C/$^{13}$C ratio and the evolution along the AGB}",
      journal = {\aap},
         year = 2014,
        month = jun,
       volume = {566},
          eid = {A145},
        pages = {A145},
          doi = {10.1051/0004-6361/201423721},
archivePrefix = {arXiv},
       eprint = {1405.6404},
 primaryClass = {astro-ph.SR},
       adsurl = {https://ui.adsabs.harvard.edu/abs/2014A&A...566A.145R}
}

@ARTICLE{bru24,
       author = {{Brunken}, N.~G.~C. and {van Dishoeck}, E.~F. and {Slavicinska}, K. and {le Gouellec}, V.~J.~M. and {Rocha}, W.~R.~M. and {Francis}, L. and {Tychoniec}, L. and {van Gelder}, M.~L. and {Navarro}, M.~G. and {Boogert}, A.~C.~A. and {Kavanagh}, P.~J. and {Nazari}, P. and {Greene}, T. and {Ressler}, M.~E. and {Majumdar}, L.},
        title = "{JOYS+ study of solid-state $^{12}$C/$^{13}$C isotope ratios in protostellar envelopes: Observations of CO and CO$_{2}$ ice with the James Webb Space Telescope}",
      journal = {\aap},
         year = 2024,
        month = dec,
       volume = {692},
          eid = {A163},
        pages = {A163},
          doi = {10.1051/0004-6361/202451794},
archivePrefix = {arXiv},
       eprint = {2409.17237},
 primaryClass = {astro-ph.SR},
       adsurl = {https://ui.adsabs.harvard.edu/abs/2024A&A...692A.163B}
}

@ARTICLE{ale15,
       author = {{Alexander}, C.~M. O'd. and {Bowden}, R. and {Fogel}, M.~L. and {Howard}, K.~T.},
        title = "{Carbonate abundances and isotopic compositions in chondrites}",
      journal = {\maps},
         year = 2015,
        month = apr,
       volume = {50},
       number = {4},
        pages = {810-833},
          doi = {10.1111/maps.12410},
       adsurl = {https://ui.adsabs.harvard.edu/abs/2015M&PS...50..810A}
}

@ARTICLE{ale07,
       author = {{Alexander}, C.~M.~O. 'D. and {Fogel}, M. and {Yabuta}, H. and {Cody}, G.~D.},
        title = "{The origin and evolution of chondrites recorded in the elemental and isotopic compositions of their macromolecular organic matter}",
      journal = {\gca},
         year = 2007,
        month = sep,
       volume = {71},
       number = {17},
        pages = {4380-4403},
          doi = {10.1016/j.gca.2007.06.052},
       adsurl = {https://ui.adsabs.harvard.edu/abs/2007GeCoA..71.4380A}
}

@ARTICLE{rom17,
       author = {{Romano}, D. and {Matteucci}, F. and {Zhang}, Z.-Y. and {Papadopoulos}, P.~P. and {Ivison}, R.~J.},
        title = "{The evolution of CNO isotopes: a new window on cosmic star formation history and the stellar IMF in the age of ALMA}",
      journal = {\mnras},
         year = 2017,
        month = sep,
       volume = {470},
       number = {1},
        pages = {401-415},
          doi = {10.1093/mnras/stx1197},
archivePrefix = {arXiv},
       eprint = {1704.06701},
 primaryClass = {astro-ph.GA},
       adsurl = {https://ui.adsabs.harvard.edu/abs/2017MNRAS.470..401R}
}

@ARTICLE{wil94,
       author = {{Wilson}, T.~L. and {Rood}, R.},
        title = "{Abundances in the Interstellar Medium}",
      journal = {\araa},
         year = 1994,
        month = jan,
       volume = {32},
        pages = {191-226},
          doi = {10.1146/annurev.aa.32.090194.001203},
       adsurl = {https://ui.adsabs.harvard.edu/abs/1994ARA&A..32..191W}
}

@ARTICLE{vis09,
       author = {{Visser}, R. and {van Dishoeck}, E.~F. and {Black}, J.~H.},
        title = "{The photodissociation and chemistry of CO isotopologues: applications to interstellar clouds and circumstellar disks}",
      journal = {\aap},
         year = 2009,
        month = aug,
       volume = {503},
       number = {2},
        pages = {323-343},
          doi = {10.1051/0004-6361/200912129},
archivePrefix = {arXiv},
       eprint = {0906.3699},
 primaryClass = {astro-ph.GA},
       adsurl = {https://ui.adsabs.harvard.edu/abs/2009A&A...503..323V}
}

@ARTICLE{guo25,
       author = {{Guo}, Yiyang and {Zhang}, Luyao and {Feng}, Fabo and {Li}, Zhao-Yu and {Pomazan}, Anton and {Yang}, Xiaohu},
        title = "{Search for Past Stellar Encounters and the Origin of 3I/ATLAS}",
      journal = {\aj},
         year = 2025,
        month = dec,
       volume = {170},
       number = {6},
          eid = {362},
        pages = {362},
          doi = {10.3847/1538-3881/ae1833},
archivePrefix = {arXiv},
       eprint = {2509.03361},
 primaryClass = {astro-ph.SR},
       adsurl = {https://ui.adsabs.harvard.edu/abs/2025AJ....170..362G}
}

@ARTICLE{mas15,
       author = {{Masseron}, T. and {Gilmore}, G.},
        title = "{Carbon, nitrogen and {\ensuremath{\alpha}}-element abundances determine the formation sequence of the Galactic thick and thin discs}",
      journal = {\mnras},
         year = 2015,
        month = oct,
       volume = {453},
       number = {2},
        pages = {1855-1866},
          doi = {10.1093/mnras/stv1731},
archivePrefix = {arXiv},
       eprint = {1503.00537},
 primaryClass = {astro-ph.SR},
       adsurl = {https://ui.adsabs.harvard.edu/abs/2015MNRAS.453.1855M}
}

@ARTICLE{fra05,
       author = {{Fray}, N. and {B{\'e}nilan}, Y. and {Cottin}, H. and {Gazeau}, M.-C. and {Crovisier}, J.},
        title = "{The origin of the CN radical in comets: A review from observations and models}",
      journal = {\planss},
         year = 2005,
        month = oct,
       volume = {53},
       number = {12},
        pages = {1243-1262},
          doi = {10.1016/j.pss.2005.06.005},
       adsurl = {https://ui.adsabs.harvard.edu/abs/2005P&SS...53.1243F}
}

@ARTICLE{kar14,
       author = {{Karakas}, Amanda I. and {Lattanzio}, John C.},
        title = "{The Dawes Review 2: Nucleosynthesis and Stellar Yields of Low- and Intermediate-Mass Single Stars}",
      journal = {\pasa},
         year = 2014,
        month = jul,
       volume = {31},
          eid = {e030},
        pages = {e030},
          doi = {10.1017/pasa.2014.21},
archivePrefix = {arXiv},
       eprint = {1405.0062},
 primaryClass = {astro-ph.SR},
       adsurl = {https://ui.adsabs.harvard.edu/abs/2014PASA...31...30K}
}

@software{Roth2026,
       author = {{Roth}, N. X.},
        title = "{jwstComet: A spectral analysis package for Python}",
howpublished = {Zenodo},
         year = 2026,
        month = feb,
          doi = {https://doi.org/10.5281/zenodo.18700806}
}

@INPROCEEDINGS{kri23,
       author = {{Krijt}, S. and {Kama}, M. and {McClure}, M. and {Teske}, J. and {Bergin}, E.~A. and {Shorttle}, O. and {Walsh}, K.~J. and {Raymond}, S.~N.},
        title = "{Chemical Habitability: Supply and Retention of Life's Essential Elements During Planet Formation}",
    booktitle = {Protostars and Planets VII},
         year = 2023,
       editor = {{Inutsuka}, S. and {Aikawa}, Y. and {Muto}, T. and {Tomida}, K. and {Tamura}, M.},
       series = {Astronomical Society of the Pacific Conference Series},
       volume = {534},
        month = jul,
        pages = {1031},
          doi = {10.48550/arXiv.2203.10056},
archivePrefix = {arXiv},
       eprint = {2203.10056},
 primaryClass = {astro-ph.EP},
       adsurl = {https://ui.adsabs.harvard.edu/abs/2023ASPC..534.1031K}
}

@ARTICLE{zak23,
       author = {{Zakharov}, V.~V. and {Rotundi}, A. and {Bockel{\'e}e-Morvan}, D. and {Bykov}, N.~Y. and {Fulle}, M. and {Biver}, N. and {Della Corte}, V. and {Rodionov}, A.~V. and {Ivanovski}, S.~L.},
        title = "{Stationary expansion of gas mixture from a spherical source into vacuum}",
      journal = {\icarus},
         year = 2023,
        month = may,
       volume = {395},
          eid = {115453},
        pages = {115453},
          doi = {10.1016/j.icarus.2023.115453},
       adsurl = {https://ui.adsabs.harvard.edu/abs/2023Icar..39515453Z}
}

@ARTICLE{cor24,
       author = {{Cordiner}, M.~A. and {Darnell}, K. and {Bockel{\'e}e-Morvan}, D. and {Roth}, N.~X. and {Biver}, N. and {Milam}, S.~N. and {Charnley}, S.~B. and {Boissier}, J. and {Bonev}, B.~P. and {Qi}, C. and {Crovisier}, J. and {Remijan}, A.~J.},
        title = "{Evidence for Surprising Heavy Nitrogen Isotopic Enrichment in Comet 46P/Wirtanen's Hydrogen Cyanide}",
      journal = {\psj},
         year = 2024,
        month = oct,
       volume = {5},
       number = {10},
          eid = {221},
        pages = {221},
          doi = {10.3847/PSJ/ad7829},
archivePrefix = {arXiv},
       eprint = {2409.05711},
 primaryClass = {astro-ph.EP},
       adsurl = {https://ui.adsabs.harvard.edu/abs/2024PSJ.....5..221C}
}

@ARTICLE{bot20,
       author = {{Botelho}, R.~B. and {Milone}, A. de C. and {Mel{\'e}ndez}, J. and {Alves-Brito}, A. and {Spina}, L. and {Bean}, J.~L.},
        title = "{Carbon, isotopic ratio $^{12}$C/$^{13}$C, and nitrogen in solar twins: constraints for the chemical evolution of the local disc}",
      journal = {\mnras},
         year = 2020,
        month = dec,
       volume = {499},
       number = {2},
        pages = {2196-2213},
          doi = {10.1093/mnras/staa2917},
archivePrefix = {arXiv},
       eprint = {2009.09003},
 primaryClass = {astro-ph.SR},
       adsurl = {https://ui.adsabs.harvard.edu/abs/2020MNRAS.499.2196B}
}

@ARTICLE{per25,
       author = {{P{\'e}rez-Couto}, X. and {Torres}, S. and {Villaver}, E. and {Mustill}, A.~J. and {Manteiga}, M.},
        title = "{3I/ATLAS: In Search of the Witnesses to Its Voyage}",
      journal = {arXiv e-prints},
         year = 2025,
        month = sep,
          eid = {arXiv:2509.07678},
        pages = {arXiv:2509.07678},
          doi = {10.48550/arXiv.2509.07678},
archivePrefix = {arXiv},
       eprint = {2509.07678},
 primaryClass = {astro-ph.EP},
       adsurl = {https://ui.adsabs.harvard.edu/abs/2025arXiv250907678P}
}

@ARTICLE{tay25,
       author = {{Taylor}, Aster G. and {Seligman}, Darryl Z.},
        title = "{The Kinematic Age of 3I/ATLAS and Its Implications for Early Planet Formation}",
      journal = {\apjl},
         year = 2025,
        month = sep,
       volume = {990},
       number = {1},
          eid = {L14},
        pages = {L14},
          doi = {10.3847/2041-8213/adfa28},
archivePrefix = {arXiv},
       eprint = {2507.08111},
 primaryClass = {astro-ph.EP},
       adsurl = {https://ui.adsabs.harvard.edu/abs/2025ApJ...990L..14T}
}

@ARTICLE{yan23,
       author = {{Yan}, Y.~T. and {Henkel}, C. and {Kobayashi}, C. and {Menten}, K.~M. and {Gong}, Y. and {Zhang}, J.~S. and {Yu}, H.~Z. and {Yang}, K. and {Xie}, J.~J. and {Wang}, Y.~X.},
        title = "{Direct measurements of carbon and sulfur isotope ratios in the Milky Way}",
      journal = {\aap},
         year = 2023,
        month = feb,
       volume = {670},
          eid = {A98},
        pages = {A98},
          doi = {10.1051/0004-6361/202244584},
archivePrefix = {arXiv},
       eprint = {2212.03252},
 primaryClass = {astro-ph.GA},
       adsurl = {https://ui.adsabs.harvard.edu/abs/2023A&A...670A..98Y}
}

@ARTICLE{rom22,
       author = {{Romano}, Donatella},
        title = "{The evolution of CNO elements in galaxies}",
      journal = {\aapr},
         year = 2022,
        month = dec,
       volume = {30},
       number = {1},
          eid = {7},
        pages = {7},
          doi = {10.1007/s00159-022-00144-z},
archivePrefix = {arXiv},
       eprint = {2210.04350},
 primaryClass = {astro-ph.GA},
       adsurl = {https://ui.adsabs.harvard.edu/abs/2022A&ARv..30....7R}
}

@ARTICLE{gua22,
       author = {{Guadarrama}, Rodrigo and {Vorobyov}, Eduard I. and {Rab}, Christian and {G{\"u}del}, Manuel},
        title = "{The effect of metallicity on the abundances of molecules in protoplanetary disks}",
      journal = {\aap},
         year = 2022,
        month = nov,
       volume = {667},
          eid = {A28},
        pages = {A28},
          doi = {10.1051/0004-6361/202140995},
archivePrefix = {arXiv},
       eprint = {2208.09327},
 primaryClass = {astro-ph.SR},
       adsurl = {https://ui.adsabs.harvard.edu/abs/2022A&A...667A..28G}
}

@ARTICLE{jen21,
       author = {{Jensen}, S.~S. and {J{\o}rgensen}, J.~K. and {Furuya}, K. and {Haugb{\o}lle}, T. and {Aikawa}, Y.},
        title = "{Modeling chemistry during star formation: water deuteration in dynamic star-forming regions}",
      journal = {\aap},
         year = 2021,
        month = may,
       volume = {649},
          eid = {A66},
        pages = {A66},
          doi = {10.1051/0004-6361/202040196},
archivePrefix = {arXiv},
       eprint = {2103.12135},
 primaryClass = {astro-ph.GA},
       adsurl = {https://ui.adsabs.harvard.edu/abs/2021A&A...649A..66J}
}

@article{don82,
  author = {Donahue, T. M. and Hoffman, J. H. and Hodges, Jr., R. R. and Watson, A. J.},
  title = {Venus Was Wet: A Measurement of the Ratio of Deuterium to Hydrogen},
  journal = {Science},
  volume = {216},
  number = {4546},
  pages = {630--633},
  year = {1982},
  doi = {10.1126/science.216.4546.630}}

@ARTICLE{boc16,
       author = {{Bockel{\'e}e-Morvan}, Dominique and {Crovisier}, J. and {Erard}, S. and {Capaccioni}, F. and {Leyrat}, C. and {Filacchione}, G. and {Drossart}, P. and {Encrenaz}, T. and {Biver}, N. and {de Sanctis}, M.-C. and {Schmitt}, B. and {K{\"u}hrt}, E. and {Capria}, M.-T. and {Combes}, M. and {Combi}, M. and {Fougere}, N. and {Arnold}, G. and {Fink}, U. and {Ip}, W. and {Migliorini}, A. and {Piccioni}, G. and {Tozzi}, G.},
        title = "{Evolution of CO$_{2}$, CH$_{4}$, and OCS abundances relative to H$_{2}$O in the coma of comet 67P around perihelion from Rosetta/VIRTIS-H observations}",
      journal = {\mnras},
         year = 2016,
        month = nov,
       volume = {462},
        pages = {S170-S183},
          doi = {10.1093/mnras/stw2428},
archivePrefix = {arXiv},
       eprint = {1609.07252},
 primaryClass = {astro-ph.EP},
       adsurl = {https://ui.adsabs.harvard.edu/abs/2016MNRAS.462S.170B}
}

@ARTICLE{bol24,
       author = {{Boley}, Kiersten M. and {Christiansen}, Jessie L. and {Zink}, Jon and {Hardegree-Ullman}, Kevin and {Lee}, Eve J. and {Hopkins}, Philip F. and {Wang}, Ji and {Fernandes}, Rachel B. and {Bergsten}, Galen J. and {Bhure}, Sakhee},
        title = "{The First Evidence of a Host Star Metallicity Cutoff in the Formation of Super-Earth Planets}",
      journal = {\aj},
         year = 2024,
        month = sep,
       volume = {168},
       number = {3},
          eid = {128},
        pages = {128},
          doi = {10.3847/1538-3881/ad6570},
archivePrefix = {arXiv},
       eprint = {2407.13821},
 primaryClass = {astro-ph.EP},
       adsurl = {https://ui.adsabs.harvard.edu/abs/2024AJ....168..128B}
}

@ARTICLE{and24,
       author = {{Andama}, Geoffrey and {Mah}, Jingyi and {Bitsch}, Bertram},
        title = "{Which stars can form planets: Planetesimal formation at low metallicities}",
      journal = {\aap},
         year = 2024,
        month = mar,
       volume = {683},
          eid = {A118},
        pages = {A118},
          doi = {10.1051/0004-6361/202348899},
archivePrefix = {arXiv},
       eprint = {2401.16155},
 primaryClass = {astro-ph.EP},
       adsurl = {https://ui.adsabs.harvard.edu/abs/2024A&A...683A.118A}
}

@ARTICLE{lee24,
       author = {{Lee}, Seokho and {Nomura}, Hideko and {Furuya}, Kenji},
        title = "{Carbon Isotope Chemistry in Protoplanetary Disks: Effects of C/O Ratios}",
      journal = {\apj},
         year = 2024,
        month = jul,
       volume = {969},
       number = {1},
          eid = {41},
        pages = {41},
          doi = {10.3847/1538-4357/ad39e3},
archivePrefix = {arXiv},
       eprint = {2404.01635},
 primaryClass = {astro-ph.GA},
       adsurl = {https://ui.adsabs.harvard.edu/abs/2024ApJ...969...41L}
}

@ARTICLE{lan84,
       author = {{Langer}, W.~D. and {Graedel}, T.~E. and {Frerking}, M.~A. and {Armentrout}, P.~B.},
        title = "{Carbon and oxygen isotope fractionation in dense interstellar clouds.}",
      journal = {\apj},
         year = 1984,
        month = feb,
       volume = {277},
        pages = {581-604},
          doi = {10.1086/161730},
       adsurl = {https://ui.adsabs.harvard.edu/abs/1984ApJ...277..581L}
}

@ARTICLE{ale10,
       author = {{Alexander}, C.~M.~O. 'D. and {Newsome}, S.~D. and {Fogel}, M.~L. and {Nittler}, L.~R. and {Busemann}, H. and {Cody}, G.~D.},
        title = "{Deuterium enrichments in chondritic macromolecular material{\textemdash}Implications for the origin and evolution of organics, water and asteroids}",
      journal = {\gca},
         year = 2010,
        month = aug,
       volume = {74},
       number = {15},
        pages = {4417-4437},
          doi = {10.1016/j.gca.2010.05.005},
       adsurl = {https://ui.adsabs.harvard.edu/abs/2010GeCoA..74.4417A}
}

@ARTICLE{gan23,
       author = {{Gandhi}, Siddharth and {de Regt}, Sam and {Snellen}, Ignas and {Zhang}, Yapeng and {Rugers}, Benson and {van Leur}, Niels and {Bosschaart}, Quincy},
        title = "{JWST Measurements of $^{13}$C, $^{18}$O, and $^{17}$O in the Atmosphere of Super-Jupiter VHS 1256 b}",
      journal = {\apjl},
         year = 2023,
        month = nov,
       volume = {957},
       number = {2},
          eid = {L36},
        pages = {L36},
          doi = {10.3847/2041-8213/ad07e2},
archivePrefix = {arXiv},
       eprint = {2311.05349},
 primaryClass = {astro-ph.EP},
       adsurl = {https://ui.adsabs.harvard.edu/abs/2023ApJ...957L..36G}
}

@ARTICLE{luo24,
       author = {{Luo}, Gan and {Colzi}, Laura and {Liu}, Tie and {Bisbas}, Thomas G. and {Li}, Di and {Sun}, Yichen and {Tang}, Ningyu},
        title = "{A new measurement of the Galactic $^{12}$C/$^{13}$C gradient from sensitive HCO$^{+}$ absorption observations}",
      journal = {\aap},
         year = 2024,
        month = oct,
       volume = {690},
          eid = {A372},
        pages = {A372},
          doi = {10.1051/0004-6361/202451412},
archivePrefix = {arXiv},
       eprint = {2409.11821},
 primaryClass = {astro-ph.GA},
       adsurl = {https://ui.adsabs.harvard.edu/abs/2024A&A...690A.372L}
}

@ARTICLE{lev10,
       author = {{Levison}, Harold F. and {Duncan}, Martin J. and {Brasser}, Ramon and {Kaufmann}, David E.},
        title = "{Capture of the Sun's Oort Cloud from Stars in Its Birth Cluster}",
      journal = {Science},
         year = 2010,
        month = jun,
       volume = {329},
       number = {5988},
        pages = {187-190},
          doi = {10.1126/science.1187535},
       adsurl = {https://ui.adsabs.harvard.edu/abs/2010Sci...329..187L}
}

@inproceedings{Villanueva2025,
	Author = {{Villanueva}, G. and {Liuzzi}, G. and {Faggi}, S. and {Protopapa}, S. and {Kofman}, V. and {Fauchez}, T. J. and {Stone}, S. W. and {Mandell}, A. M.},
	Booktitle = {Fundamentals of the Planetary Spectrum Generator 2025 Edition},
	Pages = {43-63},
	Title = {Radiative Transfer Modeling},
	Volume = {1},
	Year = {2025}}

@ARTICLE{woo25,
       author = {{Woodward}, Charles E. and {Bock{\'e}lee-Morvan}, Dominique and {Harker}, David E. and {Kelley}, Michael S.~P. and {Roth}, Nathan X. and {Wooden}, Diane H. and {Milam}, Stefanie N.},
        title = "{A JWST Study of the Remarkable Oort Cloud Comet C/2017 K2 (PanSTARRS)}",
      journal = {\psj},
         year = 2025,
        month = jun,
       volume = {6},
       number = {6},
          eid = {139},
        pages = {139},
          doi = {10.3847/PSJ/add1d5},
archivePrefix = {arXiv},
       eprint = {2504.19849},
 primaryClass = {astro-ph.EP},
       adsurl = {https://ui.adsabs.harvard.edu/abs/2025PSJ.....6..139W}
}

@ARTICLE{law23,
       author = {{Law}, David R. and {E. Morrison}, Jane and {Argyriou}, Ioannis and {Patapis}, Polychronis and {{\'A}lvarez-M{\'a}rquez}, J. and {Labiano}, Alvaro and {Vandenbussche}, Bart},
        title = "{A 3D Drizzle Algorithm for JWST and Practical Application to the MIRI Medium Resolution Spectrometer}",
      journal = {\aj},
         year = 2023,
        month = aug,
       volume = {166},
       number = {2},
          eid = {45},
        pages = {45},
          doi = {10.3847/1538-3881/acdddc},
archivePrefix = {arXiv},
       eprint = {2306.05520},
 primaryClass = {astro-ph.IM},
       adsurl = {https://ui.adsabs.harvard.edu/abs/2023AJ....166...45L}
}

@ARTICLE{cor22,
       author = {{Cordiner}, M.~A. and {Coulson}, I.~M. and {Garcia-Berrios}, E. and {Qi}, C. and {Lique}, F. and {Zo{\l}towski}, M. and {de Val-Borro}, M. and {Kuan}, Y.-J. and {Ip}, W.-H. and {Mairs}, S. and {Roth}, N.~X. and {Charnley}, S.~B. and {Milam}, S.~N. and {Tseng}, W.-L. and {Chuang}, Y.-L.},
        title = "{A SUBLIME 3D Model for Cometary Coma Emission: The Hypervolatile-rich Comet C/2016 R2 (PanSTARRS)}",
      journal = {\apj},
         year = 2022,
        month = apr,
       volume = {929},
       number = {1},
          eid = {38},
        pages = {38},
          doi = {10.3847/1538-4357/ac5893},
archivePrefix = {arXiv},
       eprint = {2202.11849},
 primaryClass = {astro-ph.EP},
       adsurl = {https://ui.adsabs.harvard.edu/abs/2022ApJ...929...38C}
}

@article{casa22,
	title        = {{CASA, the Common Astronomy Software Applications for Radio Astronomy}},
	author       = {{CASA Team} and {Bean}, Ben and {Bhatnagar}, Sanjay and {Castro}, Sandra and {Donovan Meyer}, Jennifer and {Emonts}, Bjorn and {Garcia}, Enrique and {Garwood}, Robert and {Golap}, Kumar and {Gonzalez Villalba}, Justo and {Harris}, Pamela and {Hayashi}, Yohei and {Hoskins}, Josh and {Hsieh}, Mingyu and {Jagannathan}, Preshanth and {Kawasaki}, Wataru and {Keimpema}, Aard and {Kettenis}, Mark and {Lopez}, Jorge and {Marvil}, Joshua and {Masters}, Joseph and {McNichols}, Andrew and {Mehringer}, David and {Miel}, Renaud and {Moellenbrock}, George and {Montesino}, Federico and {Nakazato}, Takeshi and {Ott}, Juergen and {Petry}, Dirk and {Pokorny}, Martin and {Raba}, Ryan and {Rau}, Urvashi and {Schiebel}, Darrell and {Schweighart}, Neal and {Sekhar}, Srikrishna and {Shimada}, Kazuhiko and {Small}, Des and {Steeb}, Jan-Willem and {Sugimoto}, Kanako and {Suoranta}, Ville and {Tsutsumi}, Takahiro and {van Bemmel}, Ilse M. and {Verkouter}, Marjolein and {Wells}, Akeem and {Xiong}, Wei and {Szomoru}, Arpad and {Griffith}, Morgan and {Glendenning}, Brian and {Kern}, Jeff},
	year         = 2022,
	month        = nov,
	journal      = {\pasp},
	volume       = 134,
	number       = 1041,
	pages        = 114501,
	doi          = {10.1088/1538-3873/ac9642},
	eid          = 114501,
	archiveprefix = {arXiv},
	eprint       = {2210.02276},
	primaryclass = {astro-ph.IM},
	adsurl       = {https://ui.adsabs.harvard.edu/abs/2022PASP..134k4501C}
}

@software{bushouse25,
  author       = {Bushouse, Howard and
                  Eisenhamer, Jonathan and
                  Dencheva, Nadia and
                  Davies, James and
                  Greenfield, Perry and
                  Morrison, Jane and
                  Hodge, Phil and
                  Simon, Bernie and
                  Grumm, David and
                  Droettboom, Michael and
                  Slavich, Edward and
                  Sosey, Megan and
                  Pauly, Tyler and
                  Miller, Todd and
                  Jedrzejewski, Robert and
                  Hack, Warren and
                  Davis, David and
                  Crawford, Steven and
                  Law, David and
                  Gordon, Karl and
                  Regan, Michael and
                  Cara, Mihai and
                  MacDonald, Ken and
                  Bradley, Larry and
                  Shanahan, Clare and
                  Jamieson, William and
                  Teodoro, Mairan and
                  Williams, Thomas and
                  Pena-Guerrero, Maria and
                  Graham, Brett and
                  Molter, Edward and
                  Brandt, Timothy and
                  Hayes, Christian and
                  Cooper, Rachel and
                  Clarke, Melanie and
                  Filippazzo, Joseph},
  title        = {JWST Calibration Pipeline},
  month        = jul,
  year         = 2025,
  publisher    = {Zenodo},
  version      = {1.19.1},
  doi          = {10.5281/zenodo.16280965},
  url          = {https://doi.org/10.5281/zenodo.16280965},
  swhid        = {swh:1:dir:c30475a0905153d5615cc51f487295e97858d972
                   ;origin=https://doi.org/10.5281/zenodo.6984365;vis
                   it=swh:1:snp:fb22ed1e75827ed3da5e29b4f7ef66f17c08c
                   7f1;anchor=swh:1:rel:7f341a76abaaf26b9b9606cb3c382
                   de28f947271;path=spacetelescope-jwst-90b5024
                  },
}

@ARTICLE{villanueva18,
       author = {{Villanueva}, G.~L. and {Smith}, M.~D. and {Protopapa}, S. and {Faggi}, S. and {Mandell}, A.~M.},
        title = "{Planetary Spectrum Generator: An accurate online radiative transfer suite for atmospheres, comets, small bodies and exoplanets}",
      journal = {\jqsrt},
         year = 2018,
        month = sep,
       volume = {217},
        pages = {86-104},
          doi = {10.1016/j.jqsrt.2018.05.023},
archivePrefix = {arXiv},
       eprint = {1803.02008},
 primaryClass = {astro-ph.EP},
       adsurl = {https://ui.adsabs.harvard.edu/abs/2018JQSRT.217...86V}
}

@INCOLLECTION{don04,
       author = {{Dones}, L. and {Weissman}, P.~R. and {Levison}, H.~F. and {Duncan}, M.~J.},
        title = "{Oort cloud formation and dynamics}",
    booktitle = {Comets II},
         year = 2004,
       editor = {{Festou}, Michel C. and {Keller}, H. Uwe and {Weaver}, Harold A.},
        pages = {153},
       adsurl = {https://ui.adsabs.harvard.edu/abs/2004come.book..153D}
}

@ARTICLE{zhe25,
       author = {{Zheng}, Xi-Ling and {Zhou}, Ji-Lin},
        title = "{Configuration of single giant planet systems generating 'Oumuamua-like interstellar asteroids}",
      journal = {\mnras},
         year = 2025,
        month = mar,
       volume = {537},
       number = {4},
        pages = {3123-3133},
          doi = {10.1093/mnras/staf234},
archivePrefix = {arXiv},
       eprint = {2502.03336},
 primaryClass = {astro-ph.EP},
       adsurl = {https://ui.adsabs.harvard.edu/abs/2025MNRAS.537.3123Z}
}

@ARTICLE{jew23,
       author = {{Jewitt}, David and {Seligman}, Darryl Z.},
        title = "{The Interstellar Interlopers}",
      journal = {\araa},
         year = 2023,
        month = aug,
       volume = {61},
        pages = {197-236},
          doi = {10.1146/annurev-astro-071221-054221},
archivePrefix = {arXiv},
       eprint = {2209.08182},
 primaryClass = {astro-ph.EP},
       adsurl = {https://ui.adsabs.harvard.edu/abs/2023ARA&A..61..197J}
}

@ARTICLE{cor20,
       author = {{Cordiner}, M.~A. and {Milam}, S.~N. and {Biver}, N. and {Bockel{\'e}e-Morvan}, D. and {Roth}, N.~X. and {Bergin}, E.~A. and {Jehin}, E. and {Remijan}, A.~J. and {Charnley}, S.~B. and {Mumma}, M.~J. and {Boissier}, J. and {Crovisier}, J. and {Paganini}, L. and {Kuan}, Y.-J. and {Lis}, D.~C.},
        title = "{Unusually high CO abundance of the first active interstellar comet}",
      journal = {Nature Astronomy},
         year = 2020,
        month = apr,
       volume = {4},
        pages = {861-866},
          doi = {10.1038/s41550-020-1087-2},
archivePrefix = {arXiv},
       eprint = {2004.09586},
 primaryClass = {astro-ph.EP},
       adsurl = {https://ui.adsabs.harvard.edu/abs/2020NatAs...4..861C}
}

@ARTICLE{stu24,
       author = {{Sturm}, J.~A. and {McClure}, M.~K. and {Harsono}, D. and {Bergner}, J.~B. and {Dartois}, E. and {Boogert}, A.~C.~A. and {Cordiner}, M.~A. and {Drozdovskaya}, M.~N. and {Ioppolo}, S. and {Law}, C.~J. and {Lis}, D.~C. and {McGuire}, B.~A. and {Melnick}, G.~J. and {Noble}, J.~A. and {{\"O}berg}, K.~I. and {Palumbo}, M.~E. and {Pendleton}, Y.~J. and {Perotti}, G. and {Rocha}, W.~R.~M. and {Urso}, R.~G. and {van Dishoeck}, E.~F.},
        title = "{A JWST/MIRI analysis of the ice distribution and polycyclic aromatic hydrocarbon emission in the protoplanetary disk HH 48 NE}",
      journal = {\aap},
         year = 2024,
        month = sep,
       volume = {689},
          eid = {A92},
        pages = {A92},
          doi = {10.1051/0004-6361/202450865},
archivePrefix = {arXiv},
       eprint = {2407.09627},
 primaryClass = {astro-ph.EP},
       adsurl = {https://ui.adsabs.harvard.edu/abs/2024A&A...689A..92S}
}

@ARTICLE{tob23,
       author = {{Tobin}, John J. and {van't Hoff}, Merel L.~R. and {Leemker}, Margot and {van Dishoeck}, Ewine F. and {Paneque-Carre{\~n}o}, Teresa and {Furuya}, Kenji and {Harsono}, Daniel and {Persson}, Magnus V. and {Cleeves}, L. Ilsedore and {Sheehan}, Patrick D. and {Cieza}, Lucas},
        title = "{Deuterium-enriched water ties planet-forming disks to comets and protostars}",
      journal = {\nat},
         year = 2023,
        month = mar,
       volume = {615},
       number = {7951},
        pages = {227-230},
          doi = {10.1038/s41586-022-05676-z},
       adsurl = {https://ui.adsabs.harvard.edu/abs/2023Natur.615..227T}
}

@ARTICLE{sla25,
       author = {{Slavicinska}, Katerina and {Tychoniec}, {\L}ukasz and {Navarro}, Mar{\'\i}a Gabriela and {van Dishoeck}, Ewine F. and {Tobin}, John J. and {van Gelder}, Martijn L. and {Chen}, Yuan and {Boogert}, A.~C. Adwin and {Drechsler}, W. Blake and {Beuther}, Henrik and {Caratti o Garatti}, Alessio and {Megeath}, S. Thomas and {Klaassen}, Pamela and {Looney}, Leslie W. and {Kavanagh}, Patrick J. and {Brunken}, Nashanty G.~C. and {Sheehan}, Patrick and {Fischer}, William J.},
        title = "{HDO Ice Detected toward an Isolated Low-mass Protostar with JWST}",
      journal = {\apjl},
         year = 2025,
        month = jun,
       volume = {986},
       number = {2},
          eid = {L19},
        pages = {L19},
          doi = {10.3847/2041-8213/addb45},
archivePrefix = {arXiv},
       eprint = {2505.14686},
 primaryClass = {astro-ph.SR},
       adsurl = {https://ui.adsabs.harvard.edu/abs/2025ApJ...986L..19S}
}

@ARTICLE{hop25,
       author = {{Hopkins}, Matthew J. and {Dorsey}, Rosemary C. and {Forbes}, John C. and {Bannister}, Michele T. and {Lintott}, Chris J. and {Leicester}, Brayden},
        title = "{From a Different Star: 3I/ATLAS in the Context of the {\={O}}tautahi{\textendash}Oxford Interstellar Object Population Model}",
      journal = {\apjl},
         year = 2025,
        month = sep,
       volume = {990},
       number = {2},
          eid = {L30},
        pages = {L30},
          doi = {10.3847/2041-8213/adfbf4},
archivePrefix = {arXiv},
       eprint = {2507.05318},
 primaryClass = {astro-ph.EP},
       adsurl = {https://ui.adsabs.harvard.edu/abs/2025ApJ...990L..30H}
}

@ARTICLE{lyo18,
       author = {{Lyons}, James R. and {Gharib-Nezhad}, Ehsan and {Ayres}, Thomas R.},
        title = "{A light carbon isotope composition for the Sun}",
      journal = {Nature Communications},
         year = 2018,
        month = mar,
       volume = {9},
          eid = {908},
        pages = {908},
          doi = {10.1038/s41467-018-03093-3},
       adsurl = {https://ui.adsabs.harvard.edu/abs/2018NatCo...9..908L}
}

@ARTICLE{sel25,
       author = {{Seligman}, Darryl Z. and {Micheli}, Marco and {Farnocchia}, Davide and {Denneau}, Larry and {Noonan}, John W. and {Hsieh}, Henry H. and {Santana-Ros}, Toni and {Tonry}, John and {Auchettl}, Katie and {Conversi}, Luca and {Devog{\`e}le}, Maxime and {Faggioli}, Laura and {Feinstein}, Adina D. and {Fenucci}, Marco and {Ferrais}, Marin and {Frincke}, Tessa and {Gillon}, Michael and {Hainaut}, Olivier R. and {Hart}, Kyle and {Hoffman}, Andrew and {Holt}, Carrie E. and {Hoogendam}, Willem B. and {Huber}, Mark E. and {Jehin}, Emmanuel and {Kareta}, Theodore and {Keane}, Jacqueline V. and {Kelley}, Michael S.~P. and {Lister}, Tim and {Mandt}, Kathleen and {Manfroid}, Jean and {Mar{\v{c}}eta}, Du{\v{s}}an and {Meech}, Karen J. and {Amine Miftah}, Mohamed and {Morgan}, Marvin and {Oca{\~n}a}, Francisco and {Pe{\~n}a-Asensio}, Eloy and {Shappee}, Benjamin J. and {Siverd}, Robert J. and {Taylor}, Aster G. and {Tucker}, Michael A. and {Wainscoat}, Richard and {Weryk}, Robert and {Wray}, James J. and {Yaginuma}, Atsuhiro and {Yang}, Bin and {Ye}, Quanzhi and {Zhang}, Qicheng},
        title = "{Discovery and Preliminary Characterization of a Third Interstellar Object: 3I/ATLAS}",
      journal = {\apjl},
         year = 2025,
        month = aug,
       volume = {989},
       number = {2},
          eid = {L36},
        pages = {L36},
          doi = {10.3847/2041-8213/adf49a},
archivePrefix = {arXiv},
       eprint = {2507.02757},
 primaryClass = {astro-ph.EP},
       adsurl = {https://ui.adsabs.harvard.edu/abs/2025ApJ...989L..36S}
}

@ARTICLE{cro19,
       author = {{Crossfield}, I.~J.~M. and {Lothringer}, J.~D. and {Flores}, B. and {Mills}, E.~A.~C. and {Freedman}, R. and {Valverde}, J. and {Miles}, B. and {Guo}, X. and {Skemer}, A.},
        title = "{Unusual Isotopic Abundances in a Fully Convective Stellar Binary}",
      journal = {\apjl},
         year = 2019,
        month = jan,
       volume = {871},
       number = {1},
          eid = {L3},
        pages = {L3},
          doi = {10.3847/2041-8213/aaf9b6},
archivePrefix = {arXiv},
       eprint = {1901.02607},
 primaryClass = {astro-ph.SR},
       adsurl = {https://ui.adsabs.harvard.edu/abs/2019ApJ...871L...3C}
}

@ARTICLE{gal95,
       author = {{Galli}, Daniele and {Palla}, Francesco and {Ferrini}, Federico and {Penco}, Umberto},
        title = "{Galactic Evolution of D and 3He}",
      journal = {\apj},
         year = 1995,
        month = apr,
       volume = {443},
        pages = {536},
          doi = {10.1086/175546},
       adsurl = {https://ui.adsabs.harvard.edu/abs/1995ApJ...443..536G}
}

@ARTICLE{dea96,
       author = {{Dearborn}, David S.~P. and {Steigman}, Gary and {Tosi}, Monica},
        title = "{Galactic Evolution of D and 3He Including Stellar Production of 3He}",
      journal = {\apj},
         year = 1996,
        month = jul,
       volume = {465},
        pages = {887},
          doi = {10.1086/177472},
archivePrefix = {arXiv},
       eprint = {astro-ph/9601117},
 primaryClass = {astro-ph},
       adsurl = {https://ui.adsabs.harvard.edu/abs/1996ApJ...465..887D}
}

@ARTICLE{taq14,
       author = {{Taquet}, Vianney and {Charnley}, Steven B. and {Sipil{\"a}}, Olli},
        title = "{Multilayer Formation and Evaporation of Deuterated Ices in Prestellar and Protostellar Cores}",
      journal = {\apj},
         year = 2014,
        month = aug,
       volume = {791},
       number = {1},
          eid = {1},
        pages = {1},
          doi = {10.1088/0004-637X/791/1/1},
archivePrefix = {arXiv},
       eprint = {1405.3268},
 primaryClass = {astro-ph.GA},
       adsurl = {https://ui.adsabs.harvard.edu/abs/2014ApJ...791....1T}
}

@ARTICLE{jac13,
       author = {{Jacquet}, Emmanuel and {Robert}, Fran{\c{c}}ois},
        title = "{Water transport in protoplanetary disks and the hydrogen isotopic composition of chondrites}",
      journal = {\icarus},
         year = 2013,
        month = apr,
       volume = {223},
       number = {2},
        pages = {722-732},
          doi = {10.1016/j.icarus.2013.01.022},
archivePrefix = {arXiv},
       eprint = {1301.5665},
 primaryClass = {astro-ph.EP},
       adsurl = {https://ui.adsabs.harvard.edu/abs/2013Icar..223..722J}
}

@ARTICLE{pag11,
       author = {{Pagani}, Laurent and {Roueff}, Evelyne and {Lesaffre}, Pierre},
        title = "{Ortho-H$_{2}$ and the Age of Interstellar Dark Clouds}",
      journal = {\apjl},
         year = 2011,
        month = oct,
       volume = {739},
       number = {2},
          eid = {L35},
        pages = {L35},
          doi = {10.1088/2041-8205/739/2/L35},
archivePrefix = {arXiv},
       eprint = {1109.6495},
 primaryClass = {astro-ph.GA},
       adsurl = {https://ui.adsabs.harvard.edu/abs/2011ApJ...739L..35P}
}

@ARTICLE{fur16,
       author = {{Furuya}, K. and {van Dishoeck}, E.~F. and {Aikawa}, Y.},
        title = "{Reconstructing the history of water ice formation from HDO/H$_{2}$O and D$_{2}$O/HDO ratios in protostellar cores}",
      journal = {\aap},
         year = 2016,
        month = feb,
       volume = {586},
          eid = {A127},
        pages = {A127},
          doi = {10.1051/0004-6361/201527579},
archivePrefix = {arXiv},
       eprint = {1512.04291},
 primaryClass = {astro-ph.GA},
       adsurl = {https://ui.adsabs.harvard.edu/abs/2016A&A...586A.127F}
}

@ARTICLE{lee15,
       author = {{Lee}, Jeong-Eun and {Bergin}, Edwin A.},
        title = "{The D/H Ratio of Water Ice at Low Temperatures}",
      journal = {\apj},
         year = 2015,
        month = jan,
       volume = {799},
       number = {1},
          eid = {104},
        pages = {104},
          doi = {10.1088/0004-637X/799/1/104},
archivePrefix = {arXiv},
       eprint = {1411.4231},
 primaryClass = {astro-ph.GA},
       adsurl = {https://ui.adsabs.harvard.edu/abs/2015ApJ...799..104L}
}

@ARTICLE{fur17,
       author = {{Furuya}, K. and {Drozdovskaya}, M.~N. and {Visser}, R. and {van Dishoeck}, E.~F. and {Walsh}, C. and {Harsono}, D. and {Hincelin}, U. and {Taquet}, V.},
        title = "{Water delivery from cores to disks: Deuteration as a probe of the prestellar inheritance of H$_{2}$O}",
      journal = {\aap},
         year = 2017,
        month = mar,
       volume = {599},
          eid = {A40},
        pages = {A40},
          doi = {10.1051/0004-6361/201629269},
archivePrefix = {arXiv},
       eprint = {1610.07286},
 primaryClass = {astro-ph.GA},
       adsurl = {https://ui.adsabs.harvard.edu/abs/2017A&A...599A..40F}
}

@ARTICLE{kob11,
       author = {{Kobayashi}, Chiaki and {Karakas}, Amanda I. and {Umeda}, Hideyuki},
        title = "{The evolution of isotope ratios in the Milky Way Galaxy}",
      journal = {\mnras},
         year = 2011,
        month = jul,
       volume = {414},
       number = {4},
        pages = {3231-3250},
          doi = {10.1111/j.1365-2966.2011.18621.x},
archivePrefix = {arXiv},
       eprint = {1102.5312},
 primaryClass = {astro-ph.GA},
       adsurl = {https://ui.adsabs.harvard.edu/abs/2011MNRAS.414.3231K}
}

@ARTICLE{rom03,
       author = {{Romano}, Donatella and {Matteucci}, Francesca},
        title = "{Nova nucleosynthesis and Galactic evolution of the CNO isotopes}",
      journal = {\mnras},
         year = 2003,
        month = jun,
       volume = {342},
       number = {1},
        pages = {185-198},
          doi = {10.1046/j.1365-8711.2003.06526.x},
archivePrefix = {arXiv},
       eprint = {astro-ph/0302233},
 primaryClass = {astro-ph},
       adsurl = {https://ui.adsabs.harvard.edu/abs/2003MNRAS.342..185R}
}

@ARTICLE{boogert15,
       author = {{Boogert}, A.~C. Adwin and {Gerakines}, Perry A. and {Whittet}, Douglas C.~B.},
        title = "{Observations of the icy universe.}",
      journal = {\araa},
         year = 2015,
        month = aug,
       volume = {53},
        pages = {541-581},
          doi = {10.1146/annurev-astro-082214-122348},
archivePrefix = {arXiv},
       eprint = {1501.05317},
 primaryClass = {astro-ph.GA},
       adsurl = {https://ui.adsabs.harvard.edu/abs/2015ARA&A..53..541B}
}

@ARTICLE{boker22,
       author = {{B{\"o}ker}, T. and {Arribas}, S. and {L{\"u}tzgendorf}, N. and {Alves de Oliveira}, C. and {Beck}, T.~L. and {Birkmann}, S. and {Bunker}, A.~J. and {Charlot}, S. and {de Marchi}, G. and {Ferruit}, P. and {Giardino}, G. and {Jakobsen}, P. and {Kumari}, N. and {L{\'o}pez-Caniego}, M. and {Maiolino}, R. and {Manjavacas}, E. and {Marston}, A. and {Moseley}, S.~H. and {Muzerolle}, J. and {Ogle}, P. and {Pirzkal}, N. and {Rauscher}, B. and {Rawle}, T. and {Rix}, H. -W. and {Sabbi}, E. and {Sargent}, B. and {Sirianni}, M. and {te Plate}, M. and {Valenti}, J. and {Willott}, C.~J. and {Zeidler}, P.},
        title = "{The Near-Infrared Spectrograph (NIRSpec) on the James Webb Space Telescope. III. Integral-field spectroscopy}",
      journal = {\aap},
         year = 2022,
        month = may,
       volume = {661},
          eid = {A82},
        pages = {A82},
          doi = {10.1051/0004-6361/202142589},
archivePrefix = {arXiv},
       eprint = {2202.03308},
 primaryClass = {astro-ph.IM},
       adsurl = {https://ui.adsabs.harvard.edu/abs/2022A&A...661A..82B}
}

@ARTICLE{cor23,
       author = {{Cordiner}, M.~A. and {Roth}, N.~X. and {Milam}, S.~N. and {Villanueva}, G.~L. and {Bockel{\'e}e-Morvan}, D. and {Remijan}, A.~J. and {Charnley}, S.~B. and {Biver}, N. and {Lis}, D.~C. and {Qi}, C. and {Bonev}, B.~P. and {Crovisier}, J. and {Boissier}, J.},
        title = "{Gas Sources from the Coma and Nucleus of Comet 46P/Wirtanen Observed Using ALMA}",
      journal = {\apj},
         year = 2023,
        month = aug,
       volume = {953},
       number = {1},
          eid = {59},
        pages = {59},
          doi = {10.3847/1538-4357/ace0bc},
archivePrefix = {arXiv},
       eprint = {2305.04822},
 primaryClass = {astro-ph.EP},
       adsurl = {https://ui.adsabs.harvard.edu/abs/2023ApJ...953...59C}
}

@ARTICLE{cor25a,
       author = {{Cordiner}, M.~A. and {Gibb}, E.~L. and {Kisiel}, Z. and {Roth}, N.~X. and {Biver}, N. and {Bockel{\'e}e-Morvan}, D. and {Boissier}, J. and {Bonev}, B.~P. and {Charnley}, S.~B. and {Coulson}, I.~M. and {Crovisier}, J. and {Drozdovskaya}, M.~N. and {Furuya}, K. and {Jin}, M. and {Kuan}, Y.-J. and {Lippi}, M. and {Lis}, D.~C. and {Milam}, S.~N. and {Opitom}, C. and {Qi}, C. and {Remijan}, A.~J.},
        title = "{A D/H ratio consistent with Earth's water in Halley-type comet 12P from ALMA HDO mapping}",
      journal = {Nature Astronomy},
         year = 2025,
        month = aug,
       volume = {9},
        pages = {1476-1485},
          doi = {10.1038/s41550-025-02614-7},
archivePrefix = {arXiv},
       eprint = {2508.05925},
 primaryClass = {astro-ph.EP},
       adsurl = {https://ui.adsabs.harvard.edu/abs/2025NatAs...9.1476C}
}

@ARTICLE{cor25,
       author = {{Cordiner}, Martin A. and {Roth}, Nathan X. and {Kelley}, Michael S.~P. and {Bodewits}, Dennis and {Charnley}, Steven B. and {Drozdovskaya}, Maria N. and {Farnocchia}, Davide and {Micheli}, Marco and {Milam}, Stefanie N. and {Opitom}, Cyrielle and {Schwamb}, Megan E. and {Thomas}, Cristina A. and {Bagnulo}, Stefano},
        title = "{JWST Detection of a Carbon-dioxide-dominated Gas Coma Surrounding Interstellar Object 3I/ATLAS}",
      journal = {\apjl},
         year = 2025,
        month = oct,
       volume = {991},
       number = {2},
          eid = {L43},
        pages = {L43},
          doi = {10.3847/2041-8213/ae0647},
archivePrefix = {arXiv},
       eprint = {2508.18209},
 primaryClass = {astro-ph.EP},
       adsurl = {https://ui.adsabs.harvard.edu/abs/2025ApJ...991L..43C}
}

@ARTICLE{zol25,
       author = {{{\.Z}{\'o}{\l}towski}, M. and {Lique}, F. and {{\.Z}uchowski}, P. and {K{\l}os}, J. and {Loreau}, J. and {Kedziera}, D.},
        title = "{Collisional excitation of HCN by H$_{2}$O}",
      journal = {\mnras},
         year = 2025,
        month = jun,
       volume = {540},
       number = {1},
        pages = {626-632},
          doi = {10.1093/mnras/staf477},
       adsurl = {https://ui.adsabs.harvard.edu/abs/2025MNRAS.540..626Z}
}

@ARTICLE{fau20,
       author = {{Faure}, A. and {Lique}, F. and {Loreau}, J.},
        title = "{The effect of CO-H$_{2}$O collisions in the rotational excitation of cometary CO}",
      journal = {\mnras},
         year = 2020,
        month = mar,
       volume = {493},
       number = {1},
        pages = {776-782},
          doi = {10.1093/mnras/staa242},
archivePrefix = {arXiv},
       eprint = {2001.09562},
 primaryClass = {astro-ph.EP},
       adsurl = {https://ui.adsabs.harvard.edu/abs/2020MNRAS.493..776F}
}

@ARTICLE{yos22,
       author = {{Yoshida}, Tomohiro C. and {Nomura}, Hideko and {Furuya}, Kenji and {Tsukagoshi}, Takashi and {Lee}, Seokho},
        title = "{A New Method for Direct Measurement of Isotopologue Ratios in Protoplanetary Disks: A Case Study of the $^{12}$CO/$^{13}$CO Ratio in the TW Hya Disk}",
      journal = {\apj},
         year = 2022,
        month = jun,
       volume = {932},
       number = {2},
          eid = {126},
        pages = {126},
          doi = {10.3847/1538-4357/ac6efb},
archivePrefix = {arXiv},
       eprint = {2204.08330},
 primaryClass = {astro-ph.EP},
       adsurl = {https://ui.adsabs.harvard.edu/abs/2022ApJ...932..126Y}
}

@ARTICLE{qi26,
       author = {{Qi}, Chunhua and {Wilner}, David J. and {Espaillat}, Catherine C.},
        title = "{Isotopic Ratios in the Disk of HD 163296}",
      journal = {\apjl},
         year = 2026,
        month = jan,
       volume = {997},
       number = {1},
          eid = {L19},
        pages = {L19},
          doi = {10.3847/2041-8213/ae2916},
archivePrefix = {arXiv},
       eprint = {2601.02593},
 primaryClass = {astro-ph.EP},
       adsurl = {https://ui.adsabs.harvard.edu/abs/2026ApJ...997L..19Q}
}

@ARTICLE{hue15,
       author = {{Huebner}, W.~F. and {Mukherjee}, J.},
        title = "{Photoionization and photodissociation rates in solar and blackbody radiation fields}",
      journal = {\planss},
         year = 2015,
        month = feb,
       volume = {106},
        pages = {11-45},
          doi = {10.1016/j.pss.2014.11.022},
       adsurl = {https://ui.adsabs.harvard.edu/abs/2015P&SS..106...11H}
}

@article{coo16,
doi = {10.3847/0004-637X/825/1/51},
url = {https://doi.org/10.3847/0004-637X/825/1/51},
year = {2016},
month = {jun},
publisher = {The American Astronomical Society},
volume = {825},
number = {1},
pages = {51},
author = {Cook, Nathaniel V. and Ragozzine, Darin and Granvik, Mikael and Stephens, Denise C.},
title = {REALISTIC DETECTABILITY OF CLOSE INTERSTELLAR COMETS},
journal = {The Astrophysical Journal}
}

@INPROCEEDINGS{nom23,
       author = {{Nomura}, H. and {Furuya}, K. and {Cordiner}, M.~A. and {Charnley}, S.~B. and {Alexander}, C.~M. O'D. and {Nixon}, C.~A. and {Guzman}, V.~V. and {Yurimoto}, H. and {Tsukagoshi}, T. and {Iino}, T.},
        title = "{The Isotopic Links from Planet Forming Regions to the Solar System}",
    booktitle = {Protostars and Planets VII},
         year = 2023,
       editor = {{Inutsuka}, S. and {Aikawa}, Y. and {Muto}, T. and {Tomida}, K. and {Tamura}, M.},
       series = {Astronomical Society of the Pacific Conference Series},
       volume = {534},
        month = jul,
        pages = {1075},
       adsurl = {https://ui.adsabs.harvard.edu/abs/2023ASPC..534.1075N}
}

@ARTICLE{mcg89,
       author = {{McGlynn}, Thomas A. and {Chapman}, Robert D.},
        title = "{On the Nondetection of Extrasolar Comets}",
      journal = {\apjl},
         year = 1989,
        month = nov,
       volume = {346},
        pages = {L105},
          doi = {10.1086/185590},
       adsurl = {https://ui.adsabs.harvard.edu/abs/1989ApJ...346L.105M}
}

@ARTICLE{whi75,
       author = {{Whipple}, F.~L.},
        title = "{Do comets play a role in galactic chemistry and gamma -ray bursts?}",
      journal = {\aj},
         year = 1975,
        month = jul,
       volume = {80},
        pages = {525-531},
          doi = {10.1086/111775},
       adsurl = {https://ui.adsabs.harvard.edu/abs/1975AJ.....80..525W}
}

@INPROCEEDINGS{2004DPS....36.3416T,
       author = {{Tholen}, D.~J. and {Chesley}, S.~R.},
        title = "{Groundbased Ephemeris Development Effort for StarDust Target Wild 2}",
    booktitle = {AAS/Division for Planetary Sciences Meeting Abstracts \#36},
         year = 2004,
       series = {AAS/Division for Planetary Sciences Meeting Abstracts},
       volume = {36},
        month = nov,
          eid = {34.16},
        pages = {34.16},
       adsurl = {https://ui.adsabs.harvard.edu/abs/2004DPS....36.3416T}
}

@ARTICLE{2021Icar..35814276F,
       author = {{Farnocchia}, Davide and {Bellerose}, Julie and {Bhaskaran}, Shyam and {Micheli}, Marco and {Weryk}, Robert},
        title = "{High-fidelity comet 67P ephemeris and predictions based on Rosetta data}",
      journal = {\icarus},
         year = 2021,
        month = apr,
       volume = {358},
          eid = {114276},
        pages = {114276},
          doi = {10.1016/j.icarus.2020.114276},
       adsurl = {https://ui.adsabs.harvard.edu/abs/2021Icar..35814276F}
}

\end{document}